# New Perspectives on the Exoplanet Radius Gap from a *Mathematica* Tool and Visualized Water Equation of State

Li Zeng[1,2], Stein B. Jacobsen[1], Eugenia Hyung[1], Amit Levi[2], Chantanelle Nava[2], James Kirk[2], Caroline Piaulet[3], Gaia Lacedelli[4,5], Dimitar D. Sasselov[2], Michail I. Petaev[1,2], Sarah T. Stewart[6], Munazza K. Alam[2], Mercedes López-Morales[2], Mario Damasso[7], and David W. Latham[2]
[1] Department of Earth and Planetary Sciences, Harvard University, 20 Oxford Street, Cambridge, MA 02138, USA; lizeng@fas.harvard.edu, astrozeng@gmail.com
[2] Harvard-Smithsonian Center for Astrophysics, 60 Garden Street, Cambridge, MA 02138, USA
[3] Department of Physics and Institute for Research on Exoplanets, Université de Montréal, Montreal, QC, Canada
[4] Department of Physics and Astronomy, Università degli Studi di Padova, Vicolo dell'Osservatorio 3, I-35122, Padova, Italy
[5] INAF—Osservatorio Astronomico di Padova, Vicolo dell'Osservatorio 5, I-35122, Padova, Italy
[6] Department of Earth and Planetary Sciences, One Shields Avenue, University of California, Davis, CA 95616, USA
[7] INAF-Osservatorio Astrofisico di Torino, Via Osservatorio, 20, I-10025 Pino Torinese TO, Italy
*Received 2021 March 1; revised 2021 September 29; accepted 2021 October 8; published 2021 December 28*

## Abstract

Recent astronomical observations obtained with the Kepler and TESS missions and their related ground-based follow-ups revealed an abundance of exoplanets with a size intermediate between Earth and Neptune ($1\ R_\oplus \lesssim R \lesssim 4\ R_\oplus$). A low occurrence rate of planets has been identified at around twice the size of Earth ($2 \times R_\oplus$), known as the exoplanet radius gap or radius valley. We explore the geometry of this gap in the mass–radius diagram, with the help of a *Mathematica* plotting tool developed with the capability of manipulating exoplanet data in multidimensional parameter space, and with the help of visualized water equations of state in the temperature–density ($T$–$\rho$) graph and the entropy–pressure ($s$–$P$) graph. We show that the radius valley can be explained by a compositional difference between smaller, predominantly rocky planets ($<2 \times R_\oplus$) and larger planets ($>2 \times R_\oplus$) that exhibit greater compositional diversity including cosmic ices (water, ammonia, methane, etc.) and gaseous envelopes. In particular, among the larger planets ($>2 \times R_\oplus$), when viewed from the perspective of planet equilibrium temperature ($T_{eq}$), the hot ones ($T_{eq} \gtrsim 900$ K) are consistent with ice-dominated composition without significant gaseous envelopes, while the cold ones ($T_{eq} \lesssim 900$ K) have more diverse compositions, including various amounts of gaseous envelopes.

*Unified Astronomy Thesaurus concepts:* Exoplanet structure (495); Exoplanet astronomy (486); Exoplanet systems (484); Exoplanets (498); Super Earths (1655); Ocean planets (1151); Mini Neptunes (1063); Habitable planets (695); Extrasolar rocky planets (511); Extrasolar ice giants (2024)

## 1. Introduction

To clarify the terminology in this paper, we collectively call the exoplanets of 1–2 $R_\oplus$ super-Earths and the exoplanets of 2–4 $R_\oplus$ sub-Neptunes. The low occurrence rate of planets at around 1.7–2 $R_\oplus$, known as the exoplanet radius gap or exoplanet radius valley, naturally separates these two planet populations and gives rise to a bimodal distribution for the overall planet population in between $1 \times R_\oplus$ and $4 \times R_\oplus$. Earlier works (Berger et al. 2018, 2020; Fulton et al. 2017; Fulton & Petigura 2018; Ginzburg et al. 2018; Petigura et al. 2018; Van Eylen 2018; Wu 2019; Zeng et al. 2017b, 2017a, 2018, 2019) focus primarily on the statistics performed on planet radius and its correlation with host stellar parameters such as spectral type or age.

Various follow-up theoretical and observational efforts such as Venturini et al. (2020), Mousis et al. (2020), Mazevet et al. (2019), Haldemann et al. (2020), Otegi et al. (2020), Kite et al. (2019, 2020), Kite & Schaefer (2021), Alessi et al. (2020), King & Wheatley (2020), Lee & Connors (2020), Rao et al. (2021), Cloutier & Menou (2019), Cloutier et al. (2020a, 2020b), Owen & Estrada (2019), Owen & Adams (2019), Gupta & Schlichting (2018, 2019, 2021), Mordasini (2020), MacDonald (2019), Martinez et al. (2019), Hardegree-Ullman et al. (2020), Van Eylen et al. (2021), Kruijssen et al. (2020), Kane et al. (2020), Luque et al. (2020), Waalkes et al. (2020), David et al. (2020), Kreidberg et al. (2020), McDonald et al. (2021), and Nava et al. (2022) have since been conducted to investigate the nature of this radius valley and, in particular, its correlation with the presence of $H_2O$ and/or a gas envelope.

The aim of this paper is to provide a *Mathematica* tool (Appendix D), which combines the theoretical mass–radius curves calculated of both the planet core (Section 2) and planet envelope (Section 3), and with the capability of selectively visualizing exoplanets of various parameter ranges from a multidimensional parameter space of the planet and host stellar properties (Section 4).

For a planet core, we invent a new dimensionless parameter $\zeta$ to differentiate the effects on mass–radius relations among different core compositions (see Equation (10)). First, we visualize the contours of different $\zeta$-values in the three-component ternary diagram of Fe metals, silicates, and water ices, which are considered to be the primary possible components of a planet core. Then, the histogram of $\zeta$ is investigated in a subsequent mass–radius plot, in order to show again the existence of a bimodal distribution, and the possibility of it being explained by merely the core composition alone.

For a planet envelope, we invent a new dimensionless parameter $z$ to generally characterize the integral effect of a light envelope added onto a core (see Equation (17)). With certain assumptions, $z$ is linked to the envelope composition or composition distribution, envelope mass or mass fraction, and the envelope's temperature or temperature profile. First, we visualize the contours of different $z$ values in the mass–radius plot, to quantitatively investigate the effect of radius increase due to an envelope. When we match the trends of these $z$-contours to the existing exoplanet data, we reconfirm that the higher-mass planets ($\geqslant 11$ $M_\oplus$) generally possess a modest amount of gas envelope, and Uranus and Neptune in our solar system naturally fit into this category. However, we also emphasize the point that such a light gas envelope, if placed merely on top of a bare rocky core, cannot explain the two peaks of the bimodal distribution, as it requires too much fine-tuning in the amount and extent of the envelope. It also does not match the planet equilibrium temperature ($T_{eq}$) trend.

By means of this *Mathematica* tool and its capability of manipulating different ranges of exoplanet parameter space, we identify the planet equilibrium temperature ($T_{eq}$) as the key parameter that differentiates different types of planets (see also Kaltenegger 2017). Then, by manipulating the range of $T_{eq}$, we show that the exoplanet radius gap is most readily explained by a clustering of rocky exoplanets on one side of the gap versus a clustering of water worlds on the other side of the gap.

Looking into $T_{eq}$, we first divide exoplanets into two samples of hot versus cold planets, the dividing line being set at an equilibrium temperature $T_{eq} \sim 900$ K, which not only is empirical but also may entail a reasonable theoretical cause, which we shall investigate separately in Section 5 and also Section 7. Then, we further narrow down to different ranges of $T_{eq}$ in order to understand different features of the exoplanet population distribution and the correlation with the radius gap.

From the observer side, we include both the new planet discoveries made available by the TESS mission (Ricker et al. 2014), which samples planets orbiting around brighter host stars from all directions in the sky, and those planets from the Kepler mission (Borucki et al. 2010) and K2 mission (Howell et al. 2014) that have mass measurements. The inclusion of new observational results from TESS populates the mass–radius diagram with more planets than ever before in the relevant mass–radius regime: mass range of 1–20 $M_\oplus$ and radius range of 1–4 $R_\oplus$, because the bright host stars of TESS allow more opportunities for ground-based radial velocity follow-ups to precisely determine the planet masses—one of the necessary inputs of a mass–radius diagram.

Examples of such recent discoveries from TESS include the TOI-561 system (Lacedelli et al. 2021; Weiss et al. 2020), TOI-130 system (Sozzetti et al. 2021), and TOI-178 system (Leleu et al. 2021).

Last but not least, in this paper we briefly summarize an insight on planet internal composition from a cosmochemistry point of view (Section 6). Then, we provide visualization of the temperature–density graph ($T$–$\rho$) and the entropy–pressure graph ($s$–$P$) for water ($H_2O$) and other cosmic ices (Section 7), which allows one to grasp the physical and chemical conditions of the water-world surfaces and interiors in a general sense. We conclude with a mass–radius plot and histogram (Figure 14) summarizing our main arguments of the entire manuscript.

We hope this *Mathematica* tool can bridge over the gap between the observers and theorists, and we to keep it as simple and lucid as possible so that it can be utilized for both research and teaching.

## 2. Method for Planet Core: Geometry of the Mass–Volume Plane

### 2.1. Viewpoint from Differential Geometry: Geodesics

A spirit of differential geometry is invoked in this method (Needham 1997, 2021).

Assuming spherical symmetry of the planetary body, let us consider planet mass $M$ as the independent variable, which can be thought of as the mass enclosed within a certain shell radius $R$ or shell volume $V$.

The specific energy $u$ (specific means per unit mass) of the planetary body is a combination of its thermochemical internal energy $u_{EOS}$ and self-gravitating specific energy $u_{grav}$. We can write it as

$$u \equiv u_{EOS}(dV/dM) + u_{grav}(V, M). \quad (1)$$

The problem of solving for planetary interior structure is reduced to finding the appropriate functional form of $V(M)$ that minimizes the total energy integral $\int u \cdot dM$ of the planet, with appropriate boundary conditions and thermodynamic constraints. This is a calculus-of-variation problem. Each solution is viewed as a geodesic emanating from the origin, where its tangential slope anywhere $dV/dM$ gives the $(1/\rho)$ at that location inside the planet.

From the perspective of Hamilton–Jacobi theory, the above integral $\int u \cdot dM$ is the Hamilton's principle function $\mathcal{S}(M, V)$, which is defined as this integral from the origin to the upper limit $(M, V)$ taken along a geodesic:

$$\mathcal{S}(M, V) \equiv \int_{geodesic}^{(M,V)} u \cdot dM. \quad (2)$$

$\mathcal{S}(M, V)$ views the variable $M$ and the variable $V$ on equal footing. Calculating the variation of $\mathcal{S}(M, V)$ with respect to the endpoint $(M,V)$ gives

$$\left\{ \frac{\partial \mathcal{S}}{\partial V} = P \right. \quad (3a)$$

$$\left\{ \frac{\partial \mathcal{S}}{\partial M} = -\mathcal{H}. \right. \quad (3b)$$

Here $P$ is pressure and $\mathcal{H}$ is the Hamiltonian, and in this case, the Hamiltonian $\mathcal{H}$ equals the thermodynamic specific enthalpy $h(P)$ plus gravitational specific energy.

Then, the first-order partial-differential Hamilton–Jacobi equation (HJE) can be used to determine the geodesic:

$$-\frac{\partial \mathcal{S}(M, V)}{\partial M} = \mathcal{H}\left(M; V, \frac{\partial \mathcal{S}(M, V)}{\partial V}\right)$$
$$= h\left(\frac{\partial \mathcal{S}(M, V)}{\partial V}\right) + u_{grav}(V, M). \quad (4)$$

### 2.2. Holzapfel EOS

Wilfried B. Holzapfel has advocated for the following equation of state (EOS) and named it the second-order adapted polynomial EOS (AP2 EOS) for extreme compression (Holzapfel 2018) by defining a dimensionless variable

**Table 1**
Table of Coefficients for Three Pure Composition EOSs

| List of Coefficients for EOSs | | | |
|---|---|---|---|
| | Fe Metals | Silicates | High-pressure Ices |
| $A$ | 56 | 100 | 18 |
| $Z$ | 26 | 50 | 10 |
| $\rho_0$ (g cm$^{-3}$) | 8.2 | 4.0 | 2.1 |
| $P_{FG0}$ (GPa) | 9300 | 3200 | 1300 |
| $K_0$ (GPa) | 148 | 200 | 110 |
| $K_0'$ | 5.86 | 4.11 | 3.91 |
| $c_0$ | 3.04 | 1.667 | 1.3656 |
| $c_2$ | 1.25 | 0 | 0 |

$\eta \equiv (\rho_0/\rho)^{1/3}$ and express pressure $P$ in terms of it:

$$P(\eta) = 3 \cdot K_0 \cdot \left(\frac{1-\eta}{\eta^5}\right) \cdot (\exp[c_0 \cdot (1-\eta)]) \cdot [1 + c_2 \cdot \eta \cdot (1-\eta)], \quad (5)$$

where the coefficients $c_0$ and $c_2$ are determined by the bulk modulus $K_0$ and its pressure derivative $K_0'$,

$$\begin{cases} c_0 = -\ln\left(\frac{K_0}{P_{FG0}}\right) \end{cases} \quad (6a)$$

$$\begin{cases} c_2 = -\frac{3}{2} \cdot (K_0' - 3) - c_0 \end{cases} \quad (6b)$$

where $P_{FG0}$ is the Fermi-gas pressure at ambient (uncompressed) density,

$$P_{FG0} = \underbrace{a_0}_{0.02337 \text{ GPa} \cdot \text{nm}^5} \cdot \left(\frac{\overbrace{Z}^{\text{total number of } e^-\text{s in } V_0}}{\underbrace{V_0}_{\text{initial (uncompressed) volume}}}\right)^{5/3}$$

$$\approx 1000 \text{ GPa} \cdot \left(\frac{Z}{A} \cdot \underbrace{\text{g cc}^{-1}}_{\rho_0}\right)^{5/3}. \quad (7)$$

The bulk modulus $K_0$ is given by the effect of the outer shell $e^-$s already, while $P_{FG0}$ is given by all $e^-$s then.

AP2 EOS has the correct asymptotic behavior under extreme compression. It is what we use for the ultra-high-pressure EOSs of metals, silicates, and ices considered in this article (Table 1). Such an EOS provides a backbone so that additional effects such as temperature and envelope can be explored further on top of it.

On the other hand, the detailed EOSs of water (H$_2$O) and cosmic ices that are applicable to the relatively low pressure regimes most relevant to water worlds less than 20 $M_\oplus$ are elaborated in Appendix B (with simplified treatment of temperature effects, critical phenomenon, and smoothed phase transitions) and Appendix C (with full treatment of temperature effects, critical phenomenon, and a complete treatment of phase transitions) and visualized in a nontraditional way in a temperature–density graph and a pressure–entropy graph in Section 7.

### 2.3. Power-series Approximation of Mass–Radius Relation in log M–log R

This mass–volume relation (Figure 1), once solved, can then be converted into a mass–radius relation, and in particular in a log–log plot. Then, this log–log mass–radius relation can be generally approximated by a power series, truncating at a certain order (here the 10th order). The higher the order included, the wider the range of applicability of the power series (here we make it applicable from $10^{-5}$ to $10^{+5}$ $M_\oplus$):

$$\begin{aligned}
\log\left(\frac{R_p}{R_\oplus}\right) &= C_0 + C_1 \cdot \left[\log\left(\frac{M_p}{M_\oplus}\right)\right] \\
&+ C_2 \cdot \left[\log\left(\frac{M_p}{M_\oplus}\right)\right]^2 + C_3 \cdot \left[\log\left(\frac{M_p}{M_\oplus}\right)\right]^3 \\
&+ C_4 \cdot \left[\log\left(\frac{M_p}{M_\oplus}\right)\right]^4 + C_5 \cdot \left[\log\left(\frac{M_p}{M_\oplus}\right)\right]^5 \\
&+ C_6 \cdot \left[\log\left(\frac{M_p}{M_\oplus}\right)\right]^6 + C_7 \cdot \left[\log\left(\frac{M_p}{M_\oplus}\right)\right]^7 \\
&+ C_8 \cdot \left[\log\left(\frac{M_p}{M_\oplus}\right)\right]^8 + C_9 \cdot \left[\log\left(\frac{M_p}{M_\oplus}\right)\right]^9 \\
&+ C_{10} \cdot \left[\log\left(\frac{M_p}{M_\oplus}\right)\right]^{10} + \ldots
\end{aligned} \quad (8)$$

The coefficients $C_0$ through $C_{10}$ can be determined by fitting for a particular EOS. This power-series then allows smooth interpolation and calculation for the effect of adding an envelope. The coefficients for the mass–radius curves of planet cores made of pure Fe metals, pure silicates, and pure high-pressure ices are given in Table 2.

### 2.4. Parameter $\zeta$: Characterizing the Effect of Different Core Compositions

For 1–20 $M_\oplus$, the mass–radius relation of a planet core of uniform composition is approximately linear in log–log space with a slope of 1/4 (Valencia et al. 2007a, 2007b; Zeng et al. 2016a, 2016b, 2019)[8]:

$$(R_c/R_\oplus) \approx \zeta \cdot (M_c/M_\oplus)^{1/4}; \quad (9)$$

alternatively, $\zeta$ can be defined as

$$\zeta \equiv \frac{(R_c/R_\oplus)}{(M_c/M_\oplus)^{1/4}}. \quad (10)$$

$\zeta$ is dimensionless and on the order of unity. $\zeta$ encodes the core chemistry. For example, for an Earth-like rocky core (1/3 iron metals plus 2/3 silicates), $\zeta \approx 1$. Increasing the proportion of iron metals would make $\zeta$ smaller (Zeng et al. 2016a). On the other hand, the presence of ices in the core would make $\zeta$ larger (Zeng et al. 2019). A simple interpolation formula summarizes this relation between $\zeta$ and the mass fractions of every component, applicable to planet cores of 1–20 $M_\oplus$:

$$\begin{aligned}
\zeta \approx [&0.86 \cdot (\text{wt.\%Fe-metals}) + 1.07 \cdot (\text{wt.\% Silicates}) \\
&+ 1.55 \cdot (\text{wt.\%Ices}) - 0.14 \cdot (\text{wt.\% Ices})^2].
\end{aligned} \quad (11)$$

---

[8] http://web.gps.caltech.edu/classes/ge131/

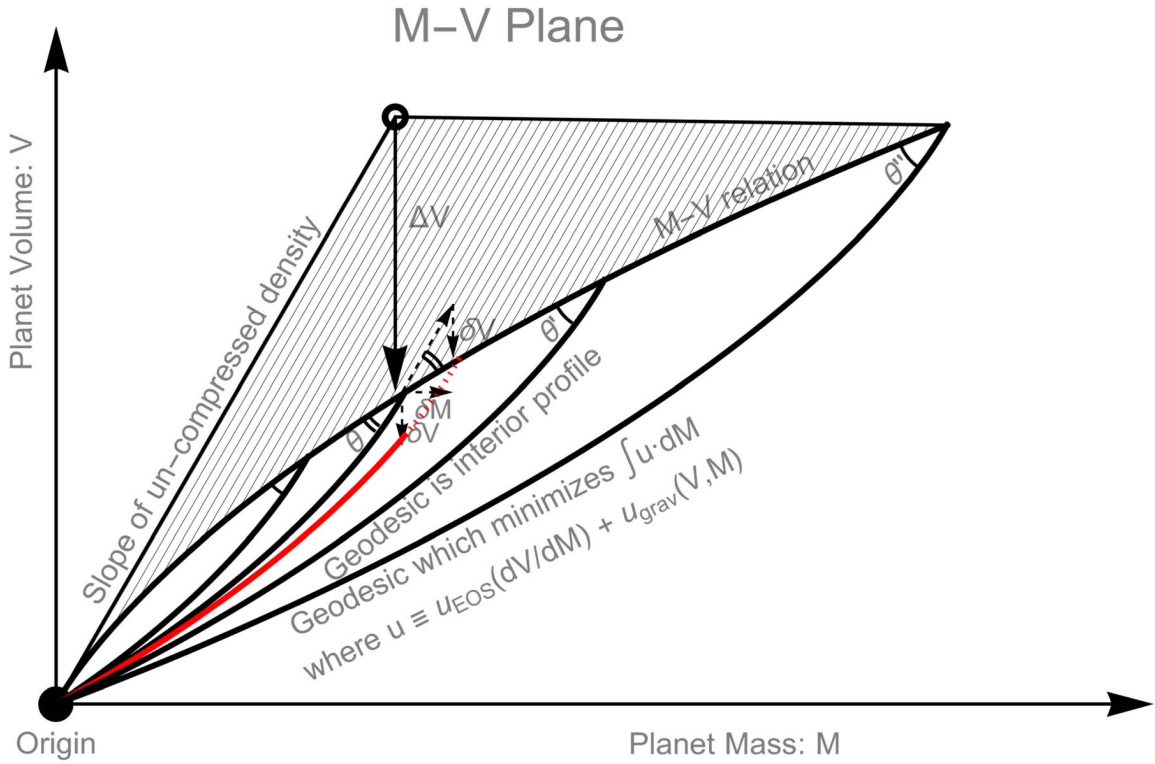

**Figure 1.** Schema of the geometry of the mass–volume plane. In particular, pay attention to the small parallelogram: $\delta V$ is the infinitesimal volume decrease of the underlying planet body owing to the addition of an infinitesimal mass increment $\delta M$ on top. This mass increment $\delta M$ pressurizes the entire underlying planet body a little bit more so that the geodesic shifts to the red curve. This mass increment $\delta M$ itself has uncompressed density and thus continues the red curve (dashed) until the M–V relation.

**Table 2**
Table of Coefficients $C_i$ for Three Pure Composition Mass–Radius Curves

| | List of Coefficients $C_i$ | | |
|---|---|---|---|
| $C_i$, where $i =$ | Fe Metals | Silicates | Ices |
| 0 | −0.11408792224566819 | 0.020013868549526272 | 0.13666292574887867 |
| 1 | 0.27851883673695 | 0.298111170324848235 | 0.27183702181443314 |
| 2 | $-1.997874049680844 \times 10^{-2}$ | $-2.012734730157388 \times 10^{-2}$ | $-7.134024332627119 \times 10^{-3}$ |
| 3 | $-2.490304269884624 \times 10^{-3}$ | $-5.2918215948260265 \times 10^{-3}$ | $-2.1407416433092126 \times 10^{-3}$ |
| 4 | $7.525048500183394 \times 10^{-5}$ | $-3.311775031243655 \times 10^{-4}$ | $-2.2608931475693915 \times 10^{-3}$ |
| 5 | $-7.162041164677924 \times 10^{-5}$ | $4.856681718363753 \times 10^{-5}$ | $-2.516518649610248 \times 10^{-4}$ |
| 6 | $-3.393158521958243 \times 10^{-5}$ | $-1.245509278944841 \times 10^{-5}$ | $1.1968169122553435 \times 10^{-4}$ |
| 7 | $8.589995554646332 \times 10^{-7}$ | $-1.3074832660503483 \times 10^{-6}$ | $1.1663496987412905 \times 10^{-5}$ |
| 8 | $1.132375249329131 \times 10^{-6}$ | $8.211419885278952 \times 10^{-7}$ | $-3.536434693875541 \times 10^{-6}$ |
| 9 | $1.132375249329131 \times 10^{-8}$ | $3.47368749025812 \times 10^{-8}$ | $-1.6848230313524644 \times 10^{-7}$ |
| 10 | $-1.0475165171649914 \times 10^{-8}$ | $-1.1251826465596989 \times 10^{-8}$ | $4.4044933682275176 \times 10^{-8}$ |

The three components always add up to unity (100%) for the planet core, and this can be seen in a ternary plot (Figure 2):

$$1 = [(\text{wt.}\%\,\text{Fe-metals}) + (\text{wt.}\%\,\text{Silicates}) + (\text{wt.}\%\,\text{Ices})]. \tag{12}$$

Equation (11) is general regardless of interior differentiation or layering of those components (Fe metals, silicated, ices) of planet cores (Zeng & Sasselov 2013). The effect of completely differentiated versus undifferentiated or partially differentiated interiors on planet radius is only on the 1% level for the mass range (1–20 $M_\oplus$) considered.

One example of the degree of differentiation is the partitioning of the Fe element between its metallic state and the oxidized state (FeO, Fe$_2$O$_3$, etc.) that would be incorporated into the silicates (Zeng & Sasselov 2013; Hyung & Jacobsen 2020). There are a few wt.% of FeO in Earth's mantle (McDonough & Sun 1995; Hyung et al. 2016).

Another example of the degree of differentiation is that the silicates may react when in contact with the supercritical fluid to become partially or wholly dissolvable in the fluid (Nisr et al. 2020; Vazan et al. 2020).

Contours of $\zeta$ according to Equation (11) can be illustrated in a ternary plot (Figure 2) to characterize the observed population separation of exoplanets. The observed exoplanet radius gap actually corresponds to the compositional gap between these two exoplanet populations (rocky vs. icy) as viewed in this ternary plot.

## 3. Method for Planet Envelope: Light Envelope

### 3.1. Parameter z: Characterizing Envelope–Core Pressure–Radius Relation

For light envelope $M_{\rm env} \lesssim 20\% \, M_c$, where $M_c$ is the planet (core) mass enclosed by the envelope, the calculation is simplified by ignoring the envelope's own effect on its gravitational field, and thus $g = (G \cdot M_c)/r^2$ within the envelope.

Hence, the hydrostatic equilibrium equation gives

$$\frac{dP}{\rho} = -g \cdot \rho = -\left(\frac{G \cdot M_c}{r^2}\right) \cdot dr = G \cdot M_c \cdot d\left(\frac{1}{r}\right); \quad (13)$$

integrating Equation (13) from envelope top to bottom, we have

$$\int_{\in {\rm envelope}} \frac{dP}{\rho} = G \cdot M_c \cdot \left(\frac{1}{R_c} - \frac{1}{R_p}\right)$$
$$= \left(\frac{G \cdot M_c}{R_c}\right) \cdot \left(1 - \frac{R_c}{R_p}\right), \quad (14)$$

where $R_c$ is the planet core radius at envelope bottom and $R_p$ is the planet radius at envelope top as defined by transit observation. Then,

$$R_p = R_c \Big/ \left[1 - \frac{\int_{\in {\rm envelope}} \frac{dP}{\rho}}{(G \cdot M_c/R_c)}\right]$$
$$= R_c \Big/ \left[1 - \frac{\int_{\in {\rm envelope}} \frac{dP}{\rho} \Big/ \left(\frac{G \cdot M_\oplus}{R_\oplus}\right)}{(M_c/M_\oplus)/(R_c/R_\oplus)}\right]. \quad (15)$$

Notice that both $\int_{\in {\rm envelope}} \frac{dP}{\rho}$ and $(G \cdot M_c/R_c)$ have the dimension of specific energy (energy per unit mass). For convenience, we normalize it by Earth units, where

$$\left(\frac{G \cdot M_\oplus}{R_\oplus}\right) = \frac{6.674 \cdot 10^{-8} \, {\rm cm}^3 \, {\rm g}^{-1} \, {\rm s}^{-2} \cdot 5.972 \cdot 10^{27} \, {\rm g}}{6.371 \cdot 10^8 \, {\rm cm}}$$
$$= 6.256 \cdot 10^{11} \, {\rm erg} \, {\rm g}^{-1}. \quad (16)$$

For convenience, let us define a new dimensionless parameter $z$:

$$z \equiv \int_{\in {\rm envelope}} \frac{dP}{\rho} \Big/ \left(\frac{G \cdot M_\oplus}{R_\oplus}\right). \quad (17)$$

Parameter $z$ can be defined generally for any envelope made up of any composition, regardless of its detailed chemistry, mean molecular weight, and temperature profile. Thus, the effect of radius boost by adding an envelope onto a planet core can be depicted by contours of $z$ as (according to Equation (15))

$$R_p = R_c \Big/ \left[1 - \frac{z}{(M_c/M_\oplus)/(R_c/R_\oplus)}\right]. \quad (18)$$

More specifically, in mass–radius plots that we generate, we choose to show the contours of $z$ of the value from 0 to 1 in incremental steps of 0.1, followed by the contours of $z$ values of 1.5, 2.0, 2.5, and 3.0.

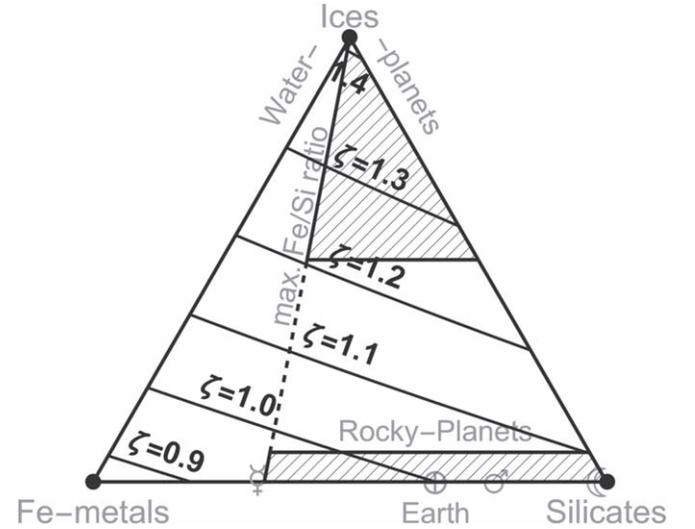

**Figure 2.** Schema of the ternary plot with contours of $\zeta \equiv (R_c/R_\oplus)/(M_c/M_\oplus)^{1/4}$. The gap is in between $\zeta = 1.1$ and $\zeta = 1.2$.

The value of $z$ depends on the envelope composition, temperature profile, and amount. The amount of envelope is directly linked to the bottom pressure of the envelope $P_{\rm env}$. In Appendix A we will provide two examples for the detailed calculation of $z$ and show how to relate it to the bottom pressure of the envelope $P_{\rm env}$.

In the next subsection, we show how to convert $P_{\rm env}$ to the envelope mass $M_{\rm env}$.

### 3.2. Envelope Pressure–Mass Relation

The hydrostatic equilibrium equation can be viewed in mass increments, in order to understand the relation between envelope mass $M_{\rm env} = (M_p - M_c)$ and envelope pressure $P_{\rm env}$ that it exerts on the planet core underneath:

$$dP = -\frac{G \cdot m \cdot dm}{4\pi \cdot r^4} = -\left(\frac{g^2}{4\pi G}\right) \cdot \frac{dm}{m}. \quad (19)$$

For a light envelope ($M_{\rm env} \lesssim 20\% \, M_c$), the integral of Equation (19) gives

$$P_{\rm env} = \int_{\in {\rm envelope}} dP \simeq \frac{\bar{g}^2}{4\pi G} \cdot \left(\frac{M_{\rm env}}{M_c}\right)$$
$$\simeq \frac{g_c^2}{4\pi G} \cdot \left(\frac{M_{\rm env}}{M_c}\right) = \left(\frac{G \cdot M_c^2}{4\pi \cdot R_c^4}\right) \cdot \left(\frac{M_{\rm env}}{M_c}\right). \quad (20)$$

Recall Equation (9), then,

$$P_{\rm env} \simeq \left(\frac{G \cdot M_\oplus^2}{4\pi \cdot R_\oplus^4}\right) \cdot \left(\frac{(M_c/M_\oplus)^2}{(R_c/R_\oplus)^4}\right) \cdot \left(\frac{M_{\rm env}/M_\oplus}{M_c/M_\oplus}\right)$$
$$\simeq (10^{11} \, {\rm Pa}) \cdot \left(\frac{1}{\zeta^4}\right) \cdot \left(\frac{M_{\rm env}}{M_\oplus}\right), \quad (21)$$

or equivalently,

$$\left(\frac{M_{\rm env}}{M_\oplus}\right) \simeq (\zeta^4) \cdot \left(\frac{P_{\rm env}}{100 \, {\rm GPa}}\right). \quad (22)$$

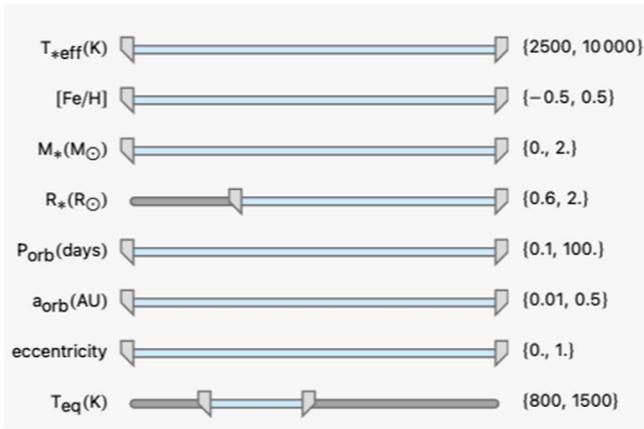

**Figure 3.** Control Pane 1 to select the intervals of eight different parameters among the host star and exoplanet properties. The eight parameters include (1) effective temperature of the host star ($T_{\star\rm eff}$), (2) metallicity as measured by iron abundance in the host star [Fe/H], (3) mass of the host star ($M_\star$) measured in solar mass ($M_\odot$), (4) radius of the host star ($R_\star$) measured in solar radius ($R_\odot$), (5) orbital period of the planet ($P_{\rm orb}$), (6) planet orbit semimajor axis ($a_{\rm orb}$), (7) orbital eccentricity of the planet ($e$), and (8) equilibrium temperature of the planet ($T_{\rm eq}$). Some of these eight parameters are interdependent on one another. However, during the selection process, we treat each parameter separately, so the overall selected parameter range is the combined proper subset of these eight intervals.

In brief, the envelope (bottom) pressure ($P_{\rm env} \equiv P_{\rm bottom}$) is roughly proportional to its mass $M_{\rm env}$. It is the pressure that the envelope exerts onto the planet core underneath. It relates to both the vertical scale ($R_p - R_c$) and the mass ($M_p - M_c$) of the envelope itself. Furthermore, it defines the physical and chemical conditions at the bottom boundary between the envelope and the planet core.

## 4. Method for Exoplanet Data Analysis from a *Mathematica* Tool

We develop a *Mathematica* tool (source code provided in Appendix D) to plot the mass–radius diagram of exoplanets.

It first draws up-to-date exoplanet data from the well-studied transiting planets of the Transiting Exoplanet Property Catalogue (TEPCat) as compiled, maintained, and updated by John Southworth at Keele University (Southworth 2011).[9]

It then selects which exoplanets to show based on the intervals of eight different parameters among the host star and exoplanet properties (Figure 3). The eight parameters include (1) effective temperature of the host star ($T_{\star\rm eff}$), (2) metallicity as measured by iron abundance in the host star [Fe/H], (3) mass of the host star ($M_\star$) measured in solar mass ($M_\odot$), (4) radius of the host star ($R_\star$) measured in solar radius ($R_\odot$), (5) orbital period of the planet ($P_{\rm orb}$), (6) planet orbit semimajor axis ($a_{\rm orb}$), (7) orbital eccentricity of the planet ($e$), and (8) equilibrium temperature of the planet ($T_{\rm eq}$). Some of these eight parameters are interdependent on one another.

Further, this tool allows one to select (1) planet mass range, (2) planet radius range, (3) planet mass error range, and (4) planet radius error range, and then either linear or logarithmic scale, for plotting (Figure 4).

Furthermore, it can show the effect of adding a gaseous envelope onto planets of different compositions (Fe metals, silicates, or ices; Figure 5).

---
[9] www.astro.keele.ac.uk/jkt/tepcat/

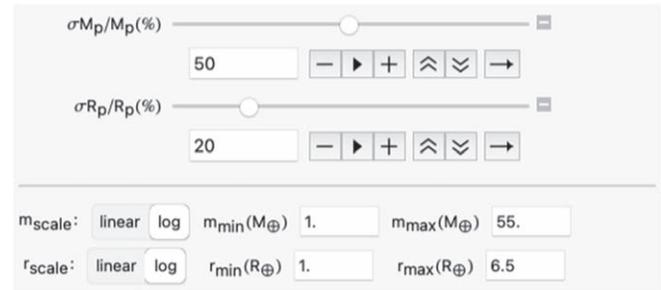

**Figure 4.** Control Pane 2 to select (1) planet mass range in Earth units, (2) planet radius range in Earth units, (3) maximum planet mass error in percentage, and (4) maximum planet radius error in percentage allowed in making the plot, and then choose either linear or logarithmic scale for plotting the *X*-axis and/or *Y*-axis.

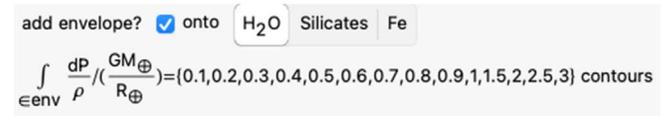

**Figure 5.** Control Pane 3 to show the general effect of adding a light gaseous envelope onto planet cores made up of different compositions (Fe metals, silicates, or ices), based on the definition of the integral parameter $z$ as in Equation (17).

In addition, this tool allows one to input new planets' data in tabulated format into the plot. Lastly, it can export the plot in PDF, EPS, or JPG format and export the properties of selected planets present in the plot in the TXT table (Figure 6).

The community is missing such an important technical tool, which we now develop and has the following advantages: (1) easy to use to make mass–radius diagrams for publication; (2) easy to expand to include other chemical species, etc.; (3) enables the easy identification of features in the distribution of the exoplanet population; and (4) open-source and freely available in Appendix D.

In brief, this *Mathematica* tool explores the different possibilities of planet compositions plus the effects of envelopes given the observables: mass, radius, $T_{\rm eq}$, etc.

## 5. Results: Mass–Radius Plots and Histograms

Here we show four figures generated by this *Mathematica* tool that can help us gain insight into the current known exoplanet population with precise mass measurements. The important parameter to separate exoplanet populations is the equilibrium temperature $T_{\rm eq}$, which is determined primarily by the amount of radiation a planet receives from its host star.

By analogy, this is similar to the broad classification of any disorder into either hot nature or cold nature in the ancient Chinese, Ayurvedic, and Hellenistic medical knowledge. This dichotomy allows us to grasp the most important factor that influences the planet populations while leaving aside other less important factors at the moment.

The plots are generated by the following selection criteria: (1) Host stars with sizes in between 0.6 and 2 $R_\odot$. (2) Planet equilibrium temperature $T_{\rm eq}$ in between 900 and 1500 K. (3) Fractional errors of planet radii less than 20% and fractional errors of planet masses less than 50%. Within the plots error bars are omitted for simplicity; however, in the *Mathematica* tool the error bars will appear as a tool tip when the mouse cursor moves over each exoplanet. (4) Solar system planets are shown for comparison as gray filled circles in the background.

**Figure 6.** Control Pane 4 to allow the import of new exoplanets' data, as well as the export of the existent plots and planets' data into various formats.

(5) The up-to-date exoplanet data are imported from TEPCat (Southworth 2011)[9]

As shown in Figure 7, the hot exoplanets (900 K ⩽ $T_{eq}$ ⩽ 1500 K) show very distinct bimodal distribution in planet radii. Two groups of planets are well separated by the radius valley. The underlying group ($\lesssim 2 \times R_\oplus$) are broadly consistent with rocky (iron metals plus silicates) composition, while the overhanging group ($\gtrsim 2 \times R_\oplus$) are broadly consistent with ice ($H_2O$, $NH_3$, $CH_4$, etc.) dominated composition. Both groups truncate sharply at ~10 $M_\oplus$. The reason for this truncation is not clear.

As shown in Figure 8, the cold exoplanets (500 K ⩽ $T_{eq}$ ⩽ 900 K) are mostly located above the pure-ice mass–radius curve. That is, they are mostly located above the radius valley. This may also be partly due to the observational biases that the small, less massive, and far-out planets, in particular, the rocky planets, are more difficult to get masses for. Excluding a few exceptions, the pure-ice mass–radius curve seems to delineate the lower bound of this population. This feature hints that their light envelopes may have derived from their ice layers. Their radius difference from the pure-ice mass–radius curve can be explained by the presence of a light envelope at most up to a few percent by mass. Also, this population of exoplanets does not truncate at ~10 $M_\oplus$, and in certain cases it can reach ~20 $M_\oplus$ or even higher masses.

In order to further our understanding of how the exoplanet distribution related to planet equilibrium temperature $T_{eq}$, we make two more plots. As shown in Figure 9, the very hot exoplanets (1500 K ⩽ $T_{eq}$ ⩽ 3000 K) are purely rocky. This is consistent with our understanding that no volatile, even ices, can exist under this condition.

As shown in Figure 10, we narrow down the $T_{eq}$ range within the hot exoplanets (1050 K ⩽ $T_{eq}$ ⩽ 1500 K). By tuning the $T_{eq}$ range, we are able to pin down a very *sharp* bimodal distribution, with each population located very tightly around either a pure-ice or pure-rocky composition mass–radius curve. The reason for this tightness may be that this $T_{eq}$ range is already hot enough to remove any gaseous envelope, while it is yet not hot enough to evaporate all the ices.

This is also a strong piece of evidence against the gas dwarf hypothesis[10] because both populations here—the exoplanet population above the radius valley and the exoplanet population below the radius valley—have almost identical (narrow) $T_{eq}$ range and mass range. It is hard to image a mechanism to keep and fine-tune the amount of a light envelope in order to fit the mass–radius distribution of exoplanet population above the radius valley to make it look like a pure-ice mass–radius curve.

In brief, our *Mathematica* tool used the observational data to find this temperature separation of 900 K empirically.

We suspect that the first possibility is that this 900 K separation has to do with a threshold of atmosphere and envelope escape and the capability of such a planet to retain water under intense radiation from the host star.

We suspect that the second possibility is that this 900 K separation has to do with the material property of water, that is, the critical point (C.P.) of water, since it defines a surrounding area within the phase diagram where sharp transitional behavior occurs from a liquid-like state to a gas-like state.

This transitional area extends up to about twice the $T_c$, about ~1300 K (Figures 11 and 12). In particular, by tracing the 0.1 GPa isobar, it inflects from a high-density liquid-like state to a low-density gas-like state.

Mendeleev views this temperature $T_c$ (C.P. temperature) as a quantitative measure of cohesion of the molecules in liquid state. When going beyond it, this cohesion is thermally diminished to zero (Karapetyants 1978).

## 6. Perspective from Cosmochemistry: Oxygen Planet

Super-Earths (1–2 $R_\oplus$) and sub-Neptunes (2–4 $R_\oplus$), regardless of their ice-rock-metal cores being dominated by silicates or ices, can be collectively called oxygen planets. When viewed from the perspective of the mass fraction, the number abundance, or the ionic volume occupied in compounds, oxygen is expected to be the dominating element.

As Table 3 shows, $^{16}_8O$ is the third most abundant element by number and by mass in the current universe, after hydrogen and helium (Anders & Grevesse 1989; Asplund et al. 2009; Lodders et al. 2009). This gross abundance pattern that results from stellar nuclear synthesis history is general throughout our own Galaxy and in other galaxies, so that the underlying stellar/planetary systems share the same basic ingredients at the current age of our universe, except some rare exceptions (Truran 1984; Truran & Heger 2003).

Also, in terms of ionic volumes in compounds, the oxygen anion $O^{2-}$ is bigger compared to cations such as $Mg^{2+}$, $Si^{4+}$, $Fe^{2+/3+}$, $Al^{3+}$, or $H^+$. Thus, the most relevant question to ask for these planets ($\lesssim 4\ R_\oplus$) is which cations are compounding with oxygen in their bulk interiors. If the dominating cations are metal-ions such as $Mg^{2+}$, $Si^{4+}$, $Fe^{2+/3+}$, and $Al^{3+}$, then this planet is a rocky planet. On the other hand, if the dominating cations are hydrogen ions $H^+$, then this planet is a water world.

---

[10] A hypothesis which explains the radius valley as a result of two groups of planets: (1) planets below the radius valley as bare rocky cores that may have lost their envelopes over time owing to certain physical mechanisms such as photoevaporation or collisions, in contrast with (2) planets above the radius valley as rocky cores that somehow retain a light hydrogen–helium-dominated envelope, aka the gas dwarfs.

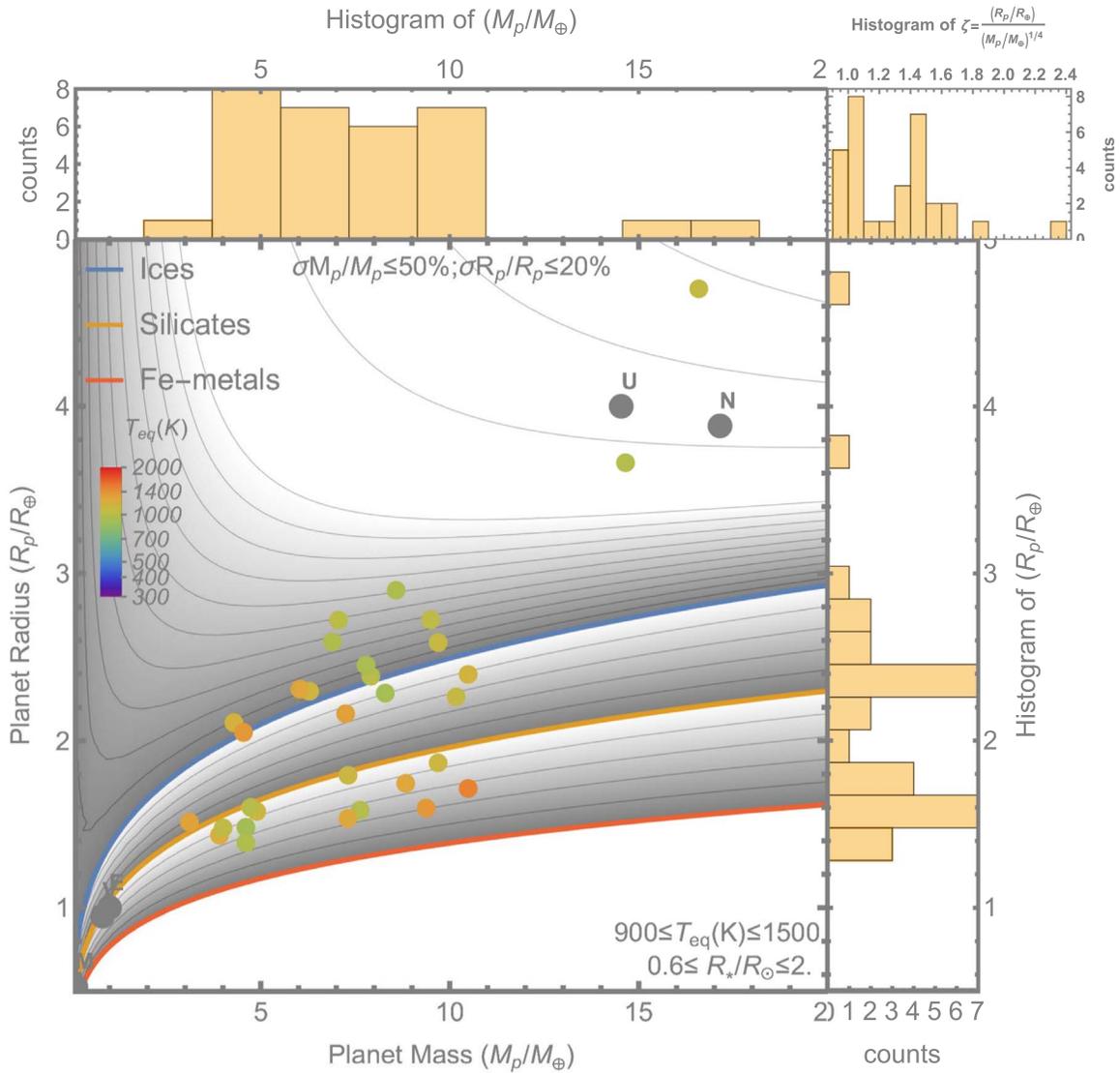

**Figure 7.** Mass–radius plot of hot (900 K ⩽ $T_{eq}$ ⩽ 1500 K) exoplanets. The bimodality of exoplanet radius distribution is distinct. The truncation of exoplanet mass distribution at ∼10 $M_\oplus$ is another prominent feature.

**Table 3**
Solar Abundance of Elements (Isotopes) (Anders & Grevesse 1989; Asplund et al. 2009; Lodders et al. 2009)

| Isotope (Normalized to Oxygen) | $^{16}_{8}O$ | $^{12}_{6}C$ | $^{14}_{7}N$ | $^{20}_{10}Ne$ | $^{28}_{14}Si$ | $^{24}_{12}Mg$ | $^{56}_{26}Fe$ | $^{32}_{16}S$ | $^{27}_{13}Al$ | $^{40}_{20}Ca$ |
|---|---|---|---|---|---|---|---|---|---|---|
| Mass fraction ratio | 1 | 0.5 | 0.2 | 0.26 | 0.1 | 0.1 | 0.2 | 0.06 | 0.01 | 0.01 |
| Atomic abundance ratio | 1 | 0.7 | 0.2 | 0.2 | 0.06 | 0.06 | 0.06 | 0.03 | 0.006 | 0.004 |

## 7. Perspective from Visualized Water Equations of State

Above we show that water worlds are abundant. In this section, with the aid of a water-ice EOS, we will describe a general picture of water-world interior structures in terms of the phases of water.

On one hand, in order to gain a clear understanding of the physics and chemistry on the surfaces and in the deep interiors of those water worlds (Levi et al. 2014, 2016, 2017; Levi & Cohen 2018; Levi & Sasselov 2018; Ramirez & Levi 2018; Yang et al. 2018; Mazevet et al. 2019; Haldemann et al. 2020; Mousis et al. 2020), one needs to delve into the details of the EOSs of the cosmic ices ($H_2O$, $NH_3$, $CH_4$, etc.) and their mixtures, in their full temperature–density–entropy–pressure parameter space, from the low-density gaseous state to the high-density condensed and compressed states, which include the fluid state and the solid state. On the other hand, this is important for constructing detailed mass–radius curves and models for planets containing these icy components and their evaporated envelopes, under various thermal, chemical, and irradiational considerations.

In particular, since $H_2O$ water is the most representative, the most well studied, and the dominating component among all the other cosmic ices, we shall start with its EOS and build our understanding of the EOS of other ices and their mixtures upon it. In many respects the EOSs of all the cosmic ices ($H_2O$, $NH_3$, $CH_4$, etc.) and their mixtures share common features, and the differences are slight modifications from one to the other. Therefore, the best approach is to visualize the EOS and see how the modifications take place.

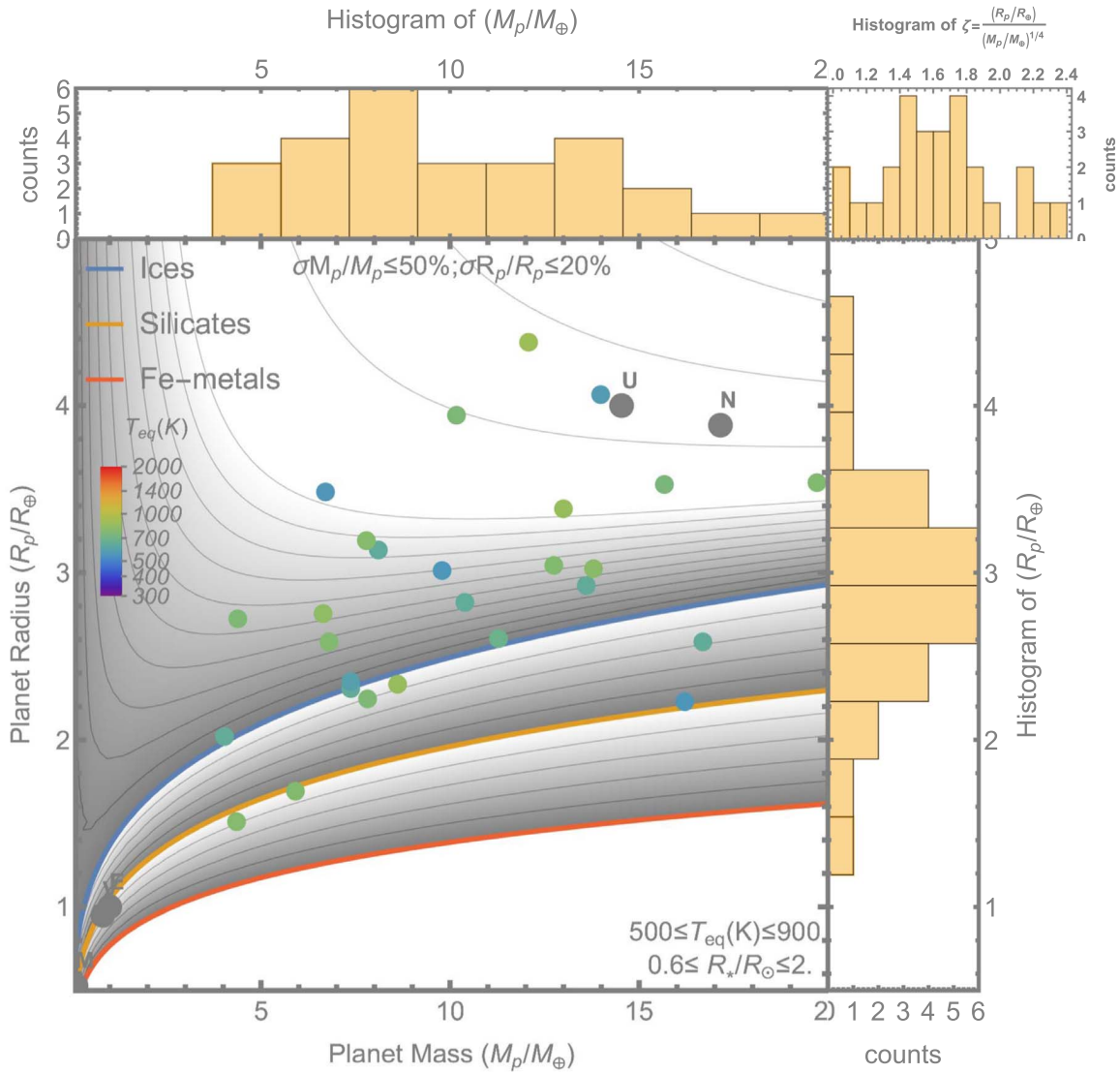

**Figure 8.** Mass–radius plot of cold (500 K ⩽ $T_{eq}$ ⩽ 900 K) exoplanets. Most of them lie above the pure-ice composition mass–radius curve. Excluding a few exceptions, it seems that the pure-ice composition mass–radius curve even delineates the lower bound of this population. This feature hints that their light envelope may have derived from their ice layers.

In Appendices B and C, we detail the $H_2O$ thermochemical data used to generate the graphs. On one hand, the pure-ice mass–radius curves in the *Mathematica* plotting tool (Figures 7 and 8) are generated with the equations in Appendix B for simplicity. The code can be expanded to include the more sophisticated $H_2O$ EOS in Appendix C; however, this requires additional assumptions about the internal temperature profile of each planet considered. The change in planet radius as a result of switching from the EOS in Appendix B to the EOS in Appendix C is very slight (percent level for temperature regimes less than 1000 K and for pressure regimes above 1 GPa; see also Figures 11 and 12, where the isotherms become more or less horizontal), which does not affect our main results and conclusions in Section 5. In contrast, the change in planet radius by adding various kinds and amounts of gaseous envelopes (the places where pressures are less than 1 GPa can be considered the envelope regime or the surface regime and can thus be included inside the integral defining parameter $z$ as in Equation (17)) is much more prominent and also sensitive to host stellar flux, compared to the thermal expansion of ices in the deep interiors. This essential idea of separating the envelope regime near the planet surface from the condensed regime of the deep interior largely simplifies our calculations of mass–radius curves without losing any generalities.

On the other hand, when we focus our discussions on some important features and the physical and chemical details of the visualized $H_2O$ EOS in the temperature–density ($T$–$\rho$) graph (Figures 11 and 12) and the entropy–pressure ($s$–$P$) graph (Figure 13), we necessarily adopt the EOS in Appendix C.

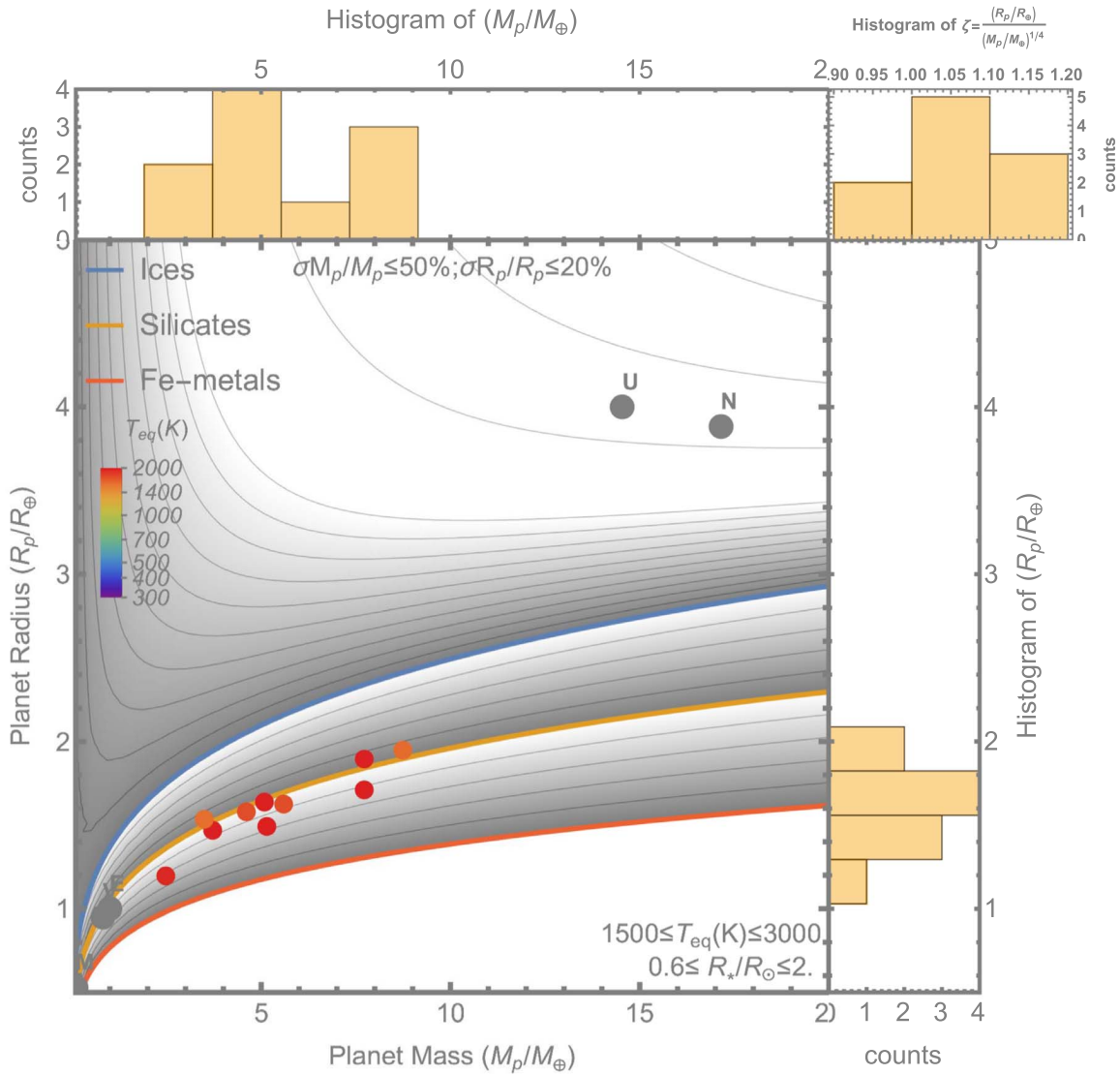

**Figure 9.** Mass–radius diagram of very hot (1500 K ⩽ $T_{eq}$ ⩽ 3000 K) exoplanets. All exoplanets here are consistent with rocky (Fe metal plus silicates) compositions. Considering their very hot nature, no volatile can exist on their surfaces.

### 7.1. Temperature–Density Graph

First, we show the $H_2O$ EOS in the temperature–density ($T$–$\rho$) graph: Figures 11 (zoomed in) and 12 (zoomed out).

The first feature of the temperature–density graph (Figures 11 and 12) is the C.P. Below the temperature of the C.P., liquid and gas phases separate and have their distinct densities. Above the temperature of the C.P., the fluid has no first-order phase transition when going from a low-density gas-like state to a high-density liquid-like state, and thus it is considered a supercritical fluid (Chaplin 2019).

The second feature of the temperature–density graph is the density discontinuity between the fluid state and the solid state along the melting curve, most prominently between Ice VII and fluid, which typically amounts to the order of a few percent (Goncharov et al. 2009; Bezacier et al. 2014; Myint et al. 2017).

The third feature is a wedge-like liquid region where low-temperature (in the vicinity of 200–300 K) liquid coexists with various ice polymorphs (Ice Ih, Ice II, Ice III, Ice V, Ice VI; Karapetyants 1978; Choukroun & Grasset 2007; Chaplin 2019;

Journaux et al. 2019, 2020). These ice polymorphs are simplified with constant densities in the graph. They occupy a relatively narrow range in log $P$ or temperature. The shape of this wedge-like liquid region can be altered by adding impurities into water due to melting-point depression. An example is shown here for adding various w.t.% of ammonia (Tillner-Roth & Friend 1998; IAPWS 2001) into water, so that the "wedge" moves toward lower temperatures. Details of the approximation are given in Appendix C.6.

The "vapor dome" (liquid–vapor transition zone with C.P. on the right-hand side in Figure 11) will retreat to lower temperatures when higher percentages of ammonia are added, which will of course warp and draw the isobars and isentropes along with it, however, only affecting the low-density ($\lesssim 1$ g cm$^{-3}$) region. Other parts of the diagram shall remain relatively intact, because the molar weights of both chemical species are very close (18 vs. 17), and they form ideal mixtures up to high pressure (Bethkenhagen et al. 2017). Liquid $NH_3$ is ~0.7 g cm$^{-3}$ (only 70% that of water) under ambient conditions, but its density will quickly catch up under pressure in both the fluid and ices. Much of the complexities only occur within a narrow region of about ∼10°–20° below 200 K (Johnson & Nicol 1987; Leliwa-Kopystyński et al. 2002; Choukroun & Grasset 2010): the hatched region in Figure 11. C.P.s of some other simple substances/compounds are given for reference (data source: Karapetyants 1978).

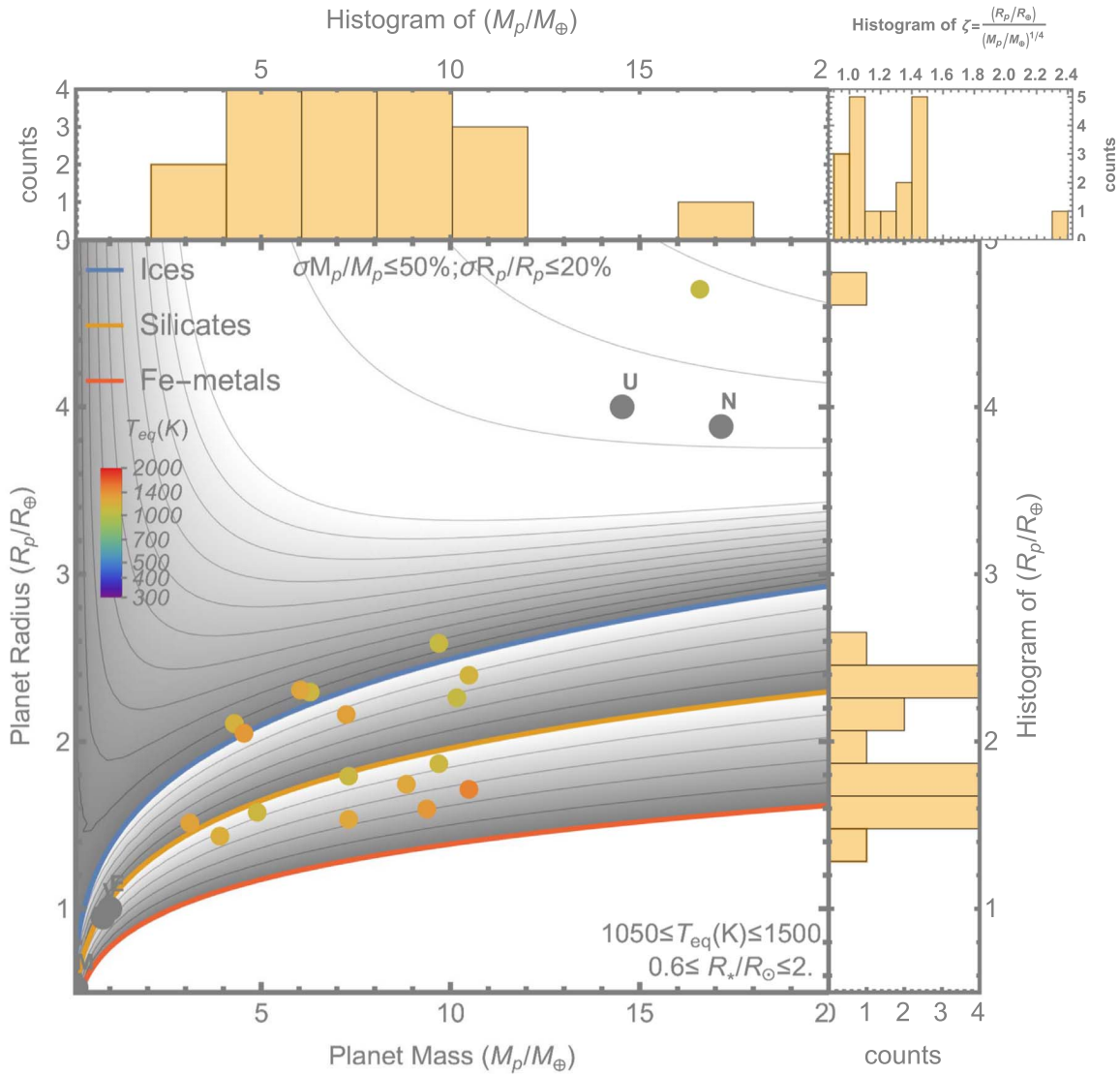

**Figure 10.** Mass–radius diagram of exoplanets with a particular $T_{eq}$ range of interest (1050 K $\leqslant T_{eq} \leqslant$ 1500 K). By tuning the $T_{eq}$ range, we are able to pin down a very sharp bimodal distribution, with each population clustering very tightly around either pure-ice or pure-rocky composition. The reason for this tightness may be that this $T_{eq}$ range is already hot enough to remove any gaseous envelope, while it is yet not hot enough to evaporate all the ices. This is also a strong piece of evidence against the gas dwarf hypothesis because both populations have almost the same $T_{eq}$.

This wedge-like liquid region is most relevant to liquid oceans on Earth and presumably other water worlds, including the interior oceans underneath ice layers on Pluto, Europa, Ganymede, etc. Because of impurities, such as various kinds of salts (e.g., NaCl) or ammonia, dissolved in them and the consequent lowering of melting points, those liquid oceans underneath ice layers must be thicker compared to pure-water ocean under the same temperature–pressure conditions.

The fourth feature is that Ice VII is the dominating solid phase of $H_2O$ from ∼2 to ∼40 GPa (Frank et al. 2004; Bezacier et al. 2014; Klotz et al. 2017). Thus, Ice VII is generally expected to be found in the deep interiors of (cold) water worlds, if the temperature is not high enough to melt it. Ice VII has (near) close-packing structure (Chaplin 2019). Ice VII can be thought of as the normal crystal structure of $H_2O$, where the effects and abnormalities of hydrogen bonding are gradually wiped off by increasing high pressure (Bridgman 1912).

In contrast, Ice Ih (hexagonal ice) that we encounter in everyday life is less dense compared to liquid water owing to loose packing caused by hydrogen bonding in the crystal. Other ice polymorphs (Ice II, Ice III, Ice V, and Ice VI) below ∼2 GPa are aligned to the left of the wedge-like liquid region, intermediate between Ice Ih and Ice VII. They are only slightly denser compared to their equilibrated liquid but less dense compared to Ice VII.

The fifth feature is the existence of a superionic state (Cavazzoni et al. 1999; Hernandez & Caracas 2016, 2018; Millot et al. 2018, 2019), where the protons ($H^+$) in the crystal are mobilized and can conduct electric current, while the oxygen anions still reside on crystal lattice points and form the backbone of the crystal structure. Thus, this state can be considered a semifluidic state where protons can be considered molten or fluidic while oxygens are not. See more details in Appendix C.3.

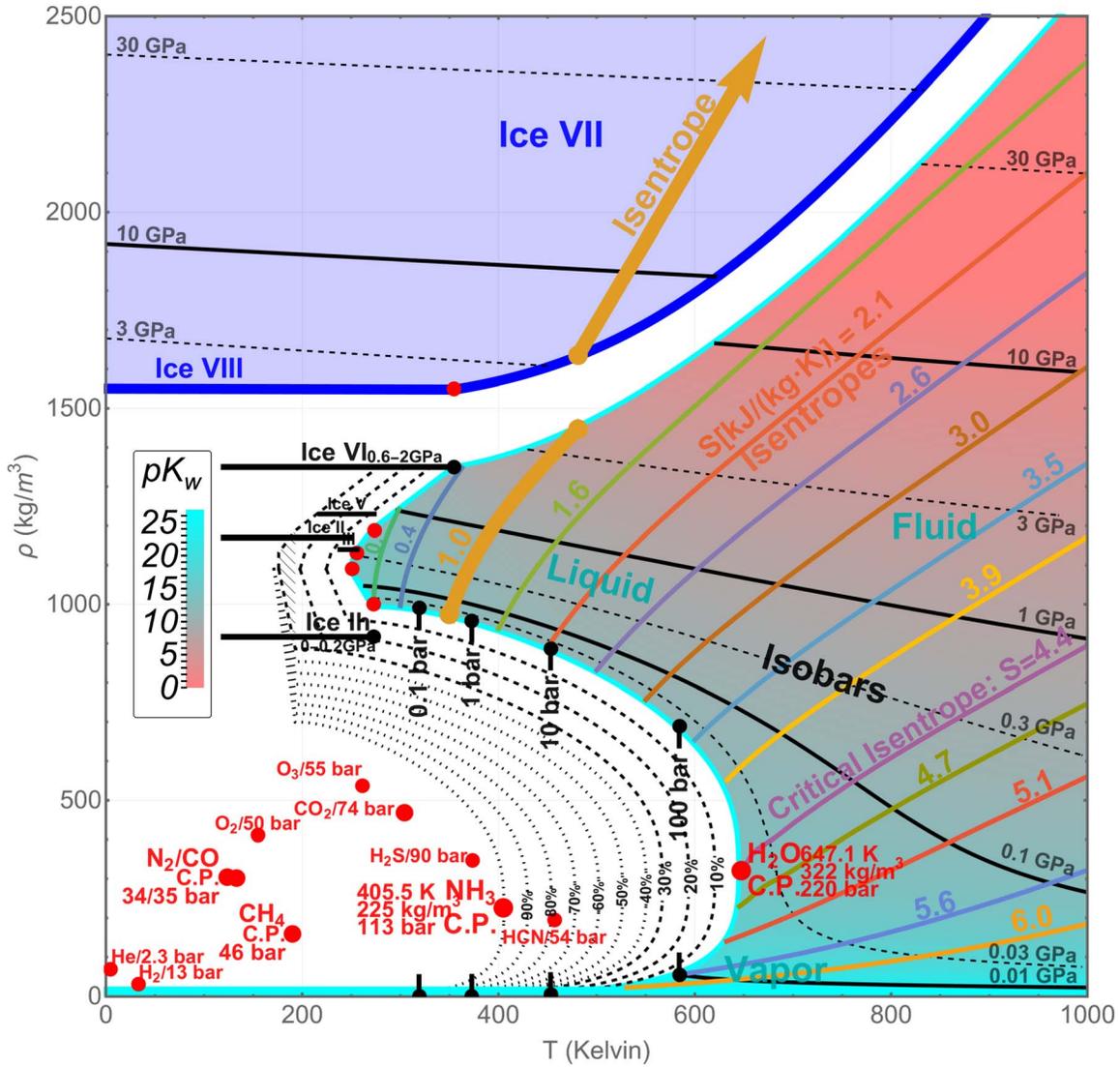

**Figure 11.** Temperature–density graph of water (zoomed in). It shows the H$_2$O phase diagram in the temperature vs. density space. The $X$-axis is the linear scale of temperature in units of kelvin. The $y$-axis is the linear scale in density in units of kg m$^{-3}$. The colored zones include (1) the fluid zone on the right that includes liquid, vapor, and supercritical fluid, (2) the extensive solid zone on the top, and (3) ice-polymorphs (Ice I(hexagonal), Ice II, Ice III, Ice V, and Ice VI) with distinct densities on the left simplified and depicted as horizontal solid black lines. Within (1) and (2), isobars (contours of constant pressures) are shown in black as labeled, and isentropes (contours of constant specific entropy) are shown in color as labeled. The background coloring of the entire fluid zone is according to the calculated ionization constant ($pK_w$). The C.P. of H$_2$O and the C.P. of some other selected simple substances are shown, in particular, ammonia NH$_3$. The gradual change of C.P. with its associated boundary of the vapor–liquid coexistence curve (vapor dome, dashed as labeled) of the H$_2$O–NH$_3$ mixture is also depicted.

Convective motions driven by internal heat sources in the planetary interiors will likely maintain their internal temperature profiles close to being isentropes in both the solid and fluid phases, except for the possibilities of a top or a bottom thermal boundary layer where thermal conduction becomes the dominant form of energy transport. Isentrope in the solid phase can be calculated with the knowledge of the Grüneisen parameter $\gamma$, which is dimensionless and typically on the order of unity. See Appendix C.2.

The sixth feature is that H$_2$O fluid ionizes and also dissociates under increasing temperatures strongly, as well as under increasing densities or pressures, as confirmed by experiments (Goncharov et al. 2005, 2009). Similarly, complex dissociation and significant increase in electrical conductivity also occur for NH$_3$ fluid in the pressure range of 10–100 GPa (Ravasio et al. 2021). The background coloring of the entire fluid region in Figures 11 and 12 depicts the ionization constant $pK_w$ of H$_2$O according to IAPWS (2019). See Appendix C.5.

### 7.2. Entropy–Pressure Graph

The isentropes and isobars in the temperature–density ($T$–$\rho$) graph become the abscissae ($x$-axis) and ordinate ($y$-axis) in the entropy–pressure ($s$–$P$) graph (Boatman 1960; Codegone 1964). The first feature of the entropy–pressure graph (Figure 13) is the "vapor dome," i.e., the liquid–vapor transition zone with C. P. on top and the triple point line at the bottom. The left branch represents "saturated liquid," and the right branch represents "saturated vapor."

Material under a shock-wave compression follows a Hugoniot: the orange curve is one example of compressing a solid-state sample (ice) starting at ambient pressure of 1 bar but a relatively cold temperature (which requires prior cooling). After the shock-wave compression, when the material sample decompresses and releases its pressure, it generally follows an isentrope, which is simply a vertical line (green arrow in Figure 13). When the isentropic line intersects the "vapor dome" from above, the relative mass proportion of liquid and vapor (i.e., gas) is determined by the lever rule with two arms being the distancings to the left (saturated liquid) branch and right (saturated vapor) branch (Kraus et al. 2011; Nellis 2017; Davies et al. 2020). Wolfram Demonstrations Project[11] has other related graphs of the H$_2$O EOS: (1) pressure–enthalpy graph (Baumann 2015), (2) temperature–entropy graph (Johnston 2016), and (3) enthalpy–entropy graph (Baumann 2017).

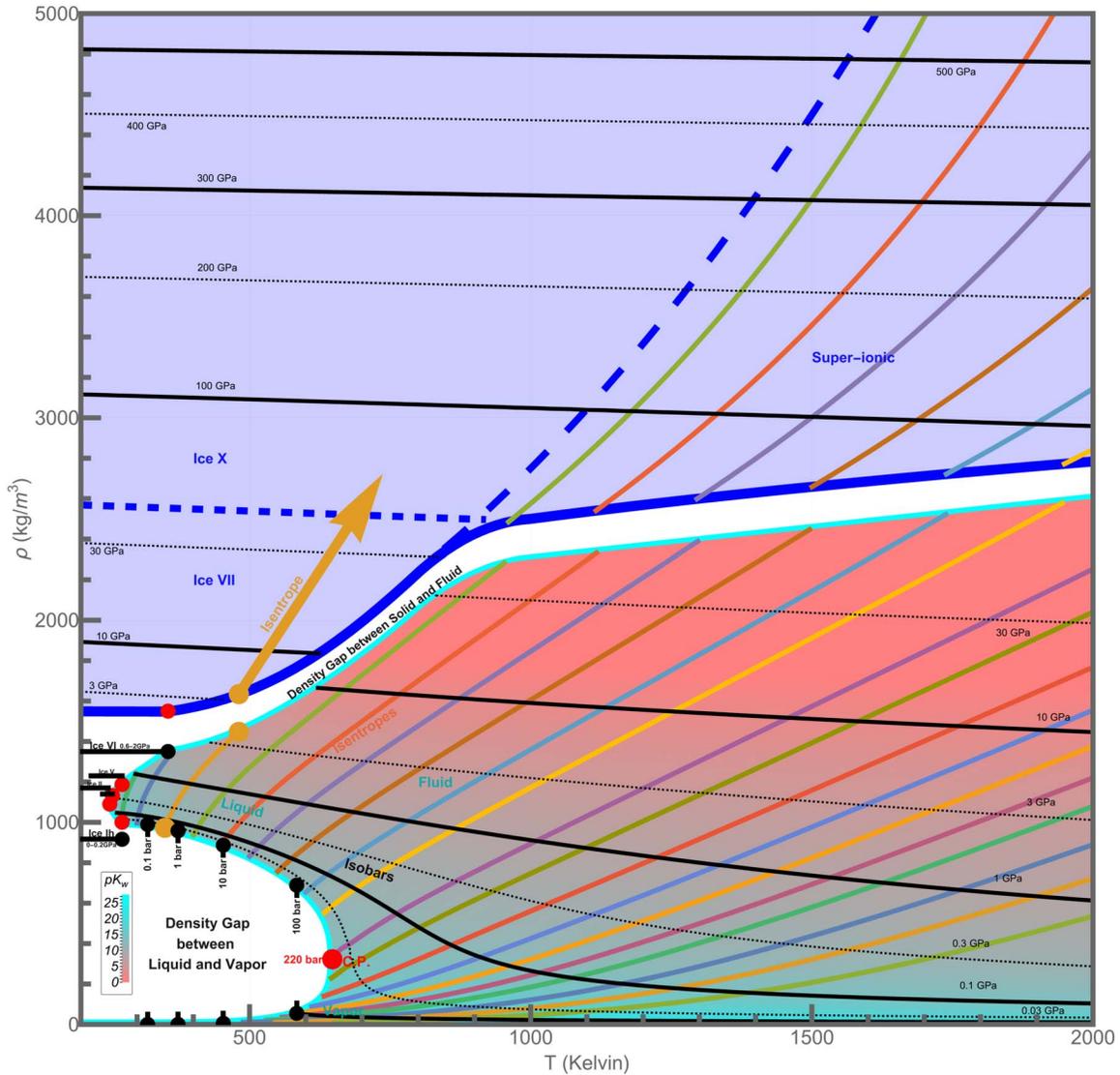

**Figure 12.** Temperature–density graph of water (zoomed out). It shows the H$_2$O phase diagram in the temperature vs. density space in a zoom-out view that includes ionic fluid and super-ionic state. X-axis is the linear scale of temperature in units of kelvin. The Y-axis is the linear scale in density in units of kg m$^{-3}$. The colored zones include (1) the fluid zone on the left that includes liquid, vapor, supercritical fluid, and ionic fluid at temperature higher than ∼1000 K; (2) the extensive solid zone on the top, which includes Ice VII, Ice X, and the super-ionic state in which hydrogen ions (H$^+$) are mobile within an oxygen crystal lattice; and (3) other ice-polymorphs ((Ice I(hexagonal), Ice II, Ice III, Ice V, and Ice VI)) with distinct densities on the left simplified and depicted as horizontal solid black lines. Within (1) and (2), isobars (contours of constant pressures) are shown in black as labeled, and isentropes (contours of constant specific entropy) are shown in color as labeled. The background coloring of the entire fluid zone is according to the calculated ionization constant ($pK_w$). The C.P. of H$_2$O and its associated boundary of the vapor–liquid coexistence curve (vapor dome, also known as the density gap between liquid and vapor) are depicted.

## 8. Conclusion

We provide a *Mathematica* tool, with built-in mass–radius curves calculated for a planet ice–rock–metal core and planet envelope, and combined with the capabilities to select and manipulate exoplanets out of a multidimensional planet and host stellar parameter space, and we plot their mass–radius diagram and histograms. In this process, we invoke two new dimensionless parameters, $\zeta$ and $z$, to characterize the effects of varying core compositions and the effects of varying envelope amount or temperature, respectively.

Through this tool, we pin down the most essential parameter that is correlated with the exoplanet radius valley, i.e., the planet equilibrium temperature ($T_{eq}$). Thus, we narrow down to various ranges of $T_{eq}$ and find that exoplanets immediately above the radius valley (within a certain equilibrium

---

[11] https://demonstrations.wolfram.com

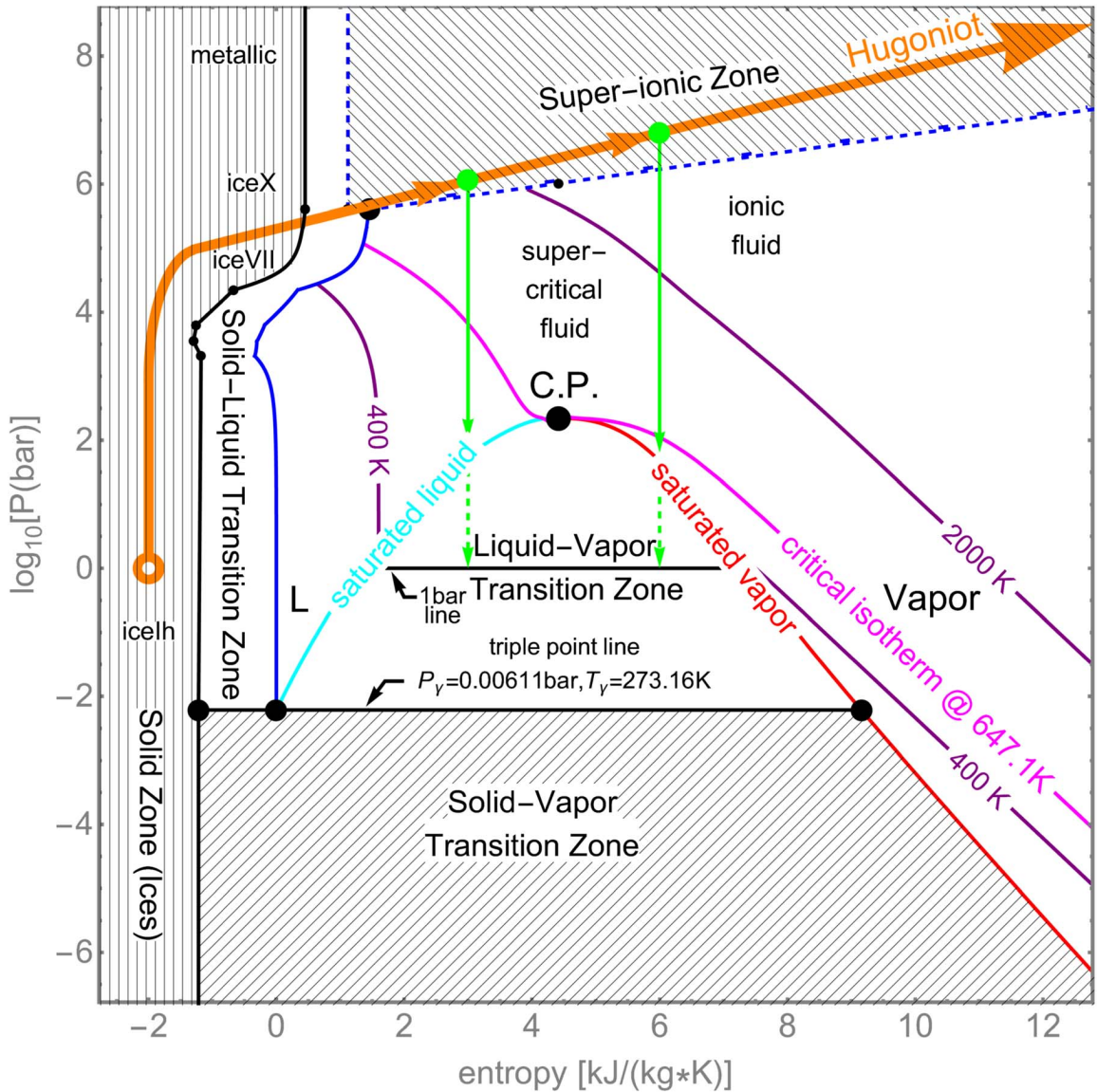

**Figure 13.** Entropy–pressure graph of water. It shows the $H_2O$ phase diagram in the entropy vs. pressure space. The X-axis is the linear scale of entropy in units of [KJ/(kg∗K)], normalized to zero at the Ice Ih–liquid–vapor triple point. The Y-axis is log base 10 of pressure in bar ($10^5$ Pa). The hatched regions include (1) solid-vapor transition zone at the bottom, (2) the solid zone (including all forms of crystal ices) to the left, and (3) the superionic zone where protons are mobile within an oxygen crystal lattice on the top. The unhatched region corresponds to the fluid, which includes both liquid and vapor, and the vapor dome, which is the coexistence region between liquid and vapor phases. Above the C.P., supercritical fluid exists, together with ionic fluid at higher temperatures. Three isotherms are shown: 400, 647.1 (at critical temperature), and 2000 K. An example of Hugoniot (shock compression curve) is shown together with two examples of subsequent pressure-release curves (green vertical downward-pointing arrows).

temperature range) are strongly consistent with water-ice-dominated composition.

We also detect a sharp truncation in the mass distribution at about 11 $M_⊕$. We circle the likely m–r parameter domain that gives rise to the two peaks. Not surprisingly, they are exactly what we had expected from our previous work (Zeng et al. 2019). The two tight clusters of exoplanets show a shift of about double the planet masses and (almost) double the planet radii: (1) a typical super-Earth is 3–5 $M_⊕$, slightly bigger than Earth, and (2) a typical water world is 6–10 $M_⊕$ and slightly bigger than 2 $R_⊕$. We summarize these main arguments in Figure 14.

We briefly touch on the cosmochemical argument that such a dichotomy of rocky versus water-rich planet cores is expected from the result of cosmic elemental abundance, which results from cosmic nuclear synthesis.

Then, in order to investigate the EOS of water and cosmic ices alike, we visualize its EOS in a nontraditional way, that is, in a temperature–density graph and a pressure–entropy graph. These two types of graphs help one to understand the different physical regimes of water and cosmic ices on such a planet's interior and surface. Especially, we choose to visualize the adiabatic profile (isentropic trajectory), isobaric profile, and Hugoniot expected in a shock-wave experiment.

Since the presence of water is the crucial ingredient for the origins of life on Earth and likely on other planets, the implication of abundant water worlds in our Galaxy and our universe suggests that life is probably a universal phenomenon, whenever and wherever other conditions, such as the availability of light/temperature, are fulfilled.

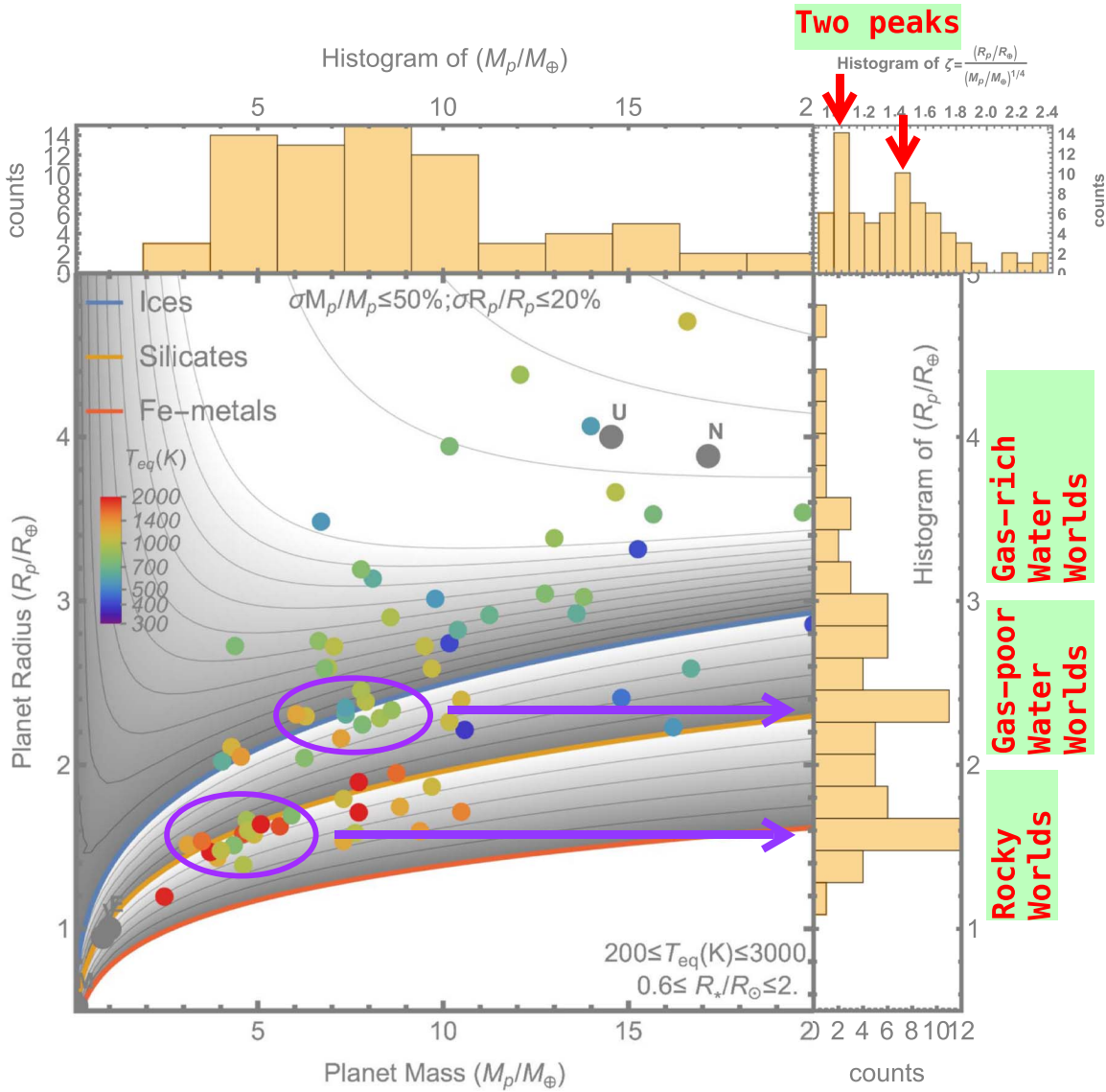

**Figure 14.** Mass–radius plot of all (200 K ⩽ $T_{eq}$ ⩽ 3000 K) small exoplanets with fractional radius errors less than 20% and fractional mass errors less than 50%. The bimodality in exoplanet radius distribution, as well as the $\zeta$-parameter, is distinct. The truncation of exoplanet mass distribution at ∼11 $M_\oplus$ is another prominent feature. We circle the likely m–r parameter domain, which gives rise to the two peaks. Not surprisingly, they are exactly what we had expected from our previous work (Zeng et al. 2019). The two tight clusters of exoplanets show a shift of about double the planet masses and (almost) double the planet radii: (1) a typical super-Earth is 3–5 $M_\oplus$, slightly bigger than Earth, and (2) a typical water world is 6–10 $M_\oplus$, slightly larger than 2.0 $R_\oplus$.

This research was partly supported by the Department of Energy National Nuclear Security Administration under awards DE-NA0003904 (to S.B.J., principal investigator) with Harvard University and by the Sandia Z Fundamental Science Program. This research represents the authors' views and not those of the Department of Energy.

This research was also partly supported by a NASA grant associated with the TESS mission. This research was also partly supported by a grant from the Simons Foundation [Simons Collaboration on the Origins of Life (Award 337090) to L.Z.]. C.P. acknowledges financial support from a scholarship awarded by the Fonds de Recherche Québécois—Nature et Technologie (FRQNT B1X and International Internships Program – Programme de stages internationaux; Québec), a Technologies for Exo-Planetary Science (TEPS) PhD scholarship and the Vanier scholarship from the Natural Sciences and Engineering Research Council (NSERC) of Canada. G.L. acknowledges support by CARIPARO Foundation, according to the agreement CARIPARO-Università degli Studi di Padova (Pratica No. 2018/0098). S.T.S. acknowledges support by NASA NExSS grant 80NSSC18K0828.

This research was also developed under the framework of the Harvard Origins of Life Initiative.

The authors would like to thank Dr. William J. Nellis from Harvard University Department of Physics for his help on the science of high-pressure shock-wave experiments; Dr. Sara Seager from Massachusetts Institute of Technology for continuous pursuits of exoplanets through the TESS mission; Dr. Douglas N. C. Lin from University of California, Santa Cruz for discussion on planet formation theory; Dr. Stephane Mazevet from Observatoire de Paris and Dr. Morris Podolak from Tel Aviv University for discussion on the $H_2O$ EOS; Dr. Mandy Bethkenhagen at Universite Lyon and Dr. Ronald

Redmer from University of Rostock for the discussion on ab initio calculations of equations of state; and Dr. Samuel Hadden, Dr. Jennifer G. Winters, and Dr. Romane Le Gal from Harvard-Smithsonian Center for Astrophysics, Dr. Aldo Bonomo and Dr. Alessandro Sozzetti from INAF-Osservatorio Astrofisico di Torino, Dr. Luca Malavolta from the University of Padua, Dr. Björn Benneke from the University of Montreal for sharing and discussing their observational data, and Dr. Jingjing Chen and Dr. David Kipping from Columbia University for discussion on the mass–radius relations.. The authors would like to thank the skillful editorial handling of the manuscript by the AAS lead editor.

*Facilities:* TESS (tess.mit.edu), Kepler (exoplanetarchive.ipac.caltech.edu/docs/KeplerMission.html), HARPS-N (https://plone.unige.ch/HARPS-N), HARPS (https://www.eso.org/public/usa/teles-instr/lasilla/36/harps/), ESPRESSO (https://www.eso.org/sci/facilities/paranal/instruments/espresso.html), CHEOPS (https://www.esa.int/Science_Exploration/Space_Science/Cheops), Gaia (https://sci.esa.int/web/gaia).

*Software:* Mathematica (Wolfram 2020; www.wolfram.com/mathematica/), TEPCat (Southworth 2011; www.astro.keele.ac.uk/jkt/tepcat/), NASA Exoplanet Archive (Akeson et al. 2013; Thompson et al. 2017; exoplanetarchive.ipac.caltech.edu).

## Appendix A
## Gaseous Envelope

Here we give two examples of calculations of the integral $\int_{\in \text{envelope}} \frac{dP}{\rho}$. As one can see from these two examples, this integral in turn relates to the bottom pressure of the envelope: $P_{\text{bottom}} \equiv P_{\text{env}}$. It is also the pressure that the envelope exerts onto the planet core underneath. Here we have made an implicit assumption that the envelope does not compress the core significantly since it is a light envelope. In the isothermal case, the top pressure $P_{\text{top}}$ of the envelope also matters when calculating this integral, and thus its vertical scale, while in the power-law case the top pressure does not matter.

### A.1. Isothermal Ideal Gas EOS

$$P_{\text{ideal}} = R \cdot T \cdot \frac{\rho}{\bar{\mu}}, \quad (A1)$$

where the mean molecular weight $\bar{\mu}$ is defined assuming volume additivity of gas species, and $X$, $Y$, $Z_i$ are mass fractions:

$$\frac{1}{\bar{\mu}} = \frac{X}{\mu_{H_2}} + \frac{Y}{\mu_{He}} + \frac{Z_1}{\mu_{Z_1}} + \frac{Z_2}{\mu_{Z_2}} + \frac{Z_3}{\mu_{Z_3}} + \ldots \quad (A2)$$

For a near cosmic mixture of hydrogen and helium ($X \approx 3/4$ and $Y \approx 1/4$), where $\bar{\mu}_{H_2-He} \approx 2.3$ and $Z \equiv \sum Z_i$, Equation (A2) can be written as

$$\frac{1}{\bar{\mu}} = \frac{1-Z}{\bar{\mu}_{H_2-He}} + \frac{Z}{\bar{\mu}_Z}. \quad (A3)$$

Furthermore, if $Z$ is not too large and $\bar{\mu}_Z \gg \bar{\mu}_{H_2-He}$, then the second term of Equation (A3) can be dropped, and we arrive at the following equation:

$$\frac{1}{\bar{\mu}} \approx \frac{1-Z}{\bar{\mu}_{H_2-He}}, \quad (A4)$$

or equivalently,

$$\bar{\mu} \approx \frac{\bar{\mu}_{H_2-He}}{(1-Z)} \quad (A5)$$

assuming $T =$ constant and $\bar{\mu} =$ constant, and introducing parameter $n$ as the number density or molar density,

$$\int_{\in \text{envelope}} \frac{dP}{\rho} = \int_{\in \text{envelope}} \frac{R \cdot T}{\bar{\mu}} \cdot \frac{d\rho}{\rho}$$
$$= \left(\frac{R \cdot T}{\bar{\mu}}\right) \cdot \ln\left(\frac{\rho_{\text{bottom}}}{\rho_{\text{top}}}\right)$$
$$= \frac{RT}{\bar{\mu}} \ln\left(\frac{n_{\text{bottom}}}{n_{\text{top}}}\right) = \frac{RT}{\bar{\mu}} \ln\left(\frac{P_{\text{bottom}}}{P_{\text{top}}}\right). \quad (A6)$$

By plugging in the number of $(G \cdot M_\oplus / R_\oplus)$ and ideal gas constant $R = 8.314 \times 10^7$ erg/(mol · K), one obtains the natural logarithm of the pressure ratio of envelope bottom over envelope top as

$$\ln\left(\frac{P_{\text{env}}}{P_{\text{top}}}\right) = 7.5 \cdot \left(\frac{\bar{\mu} \text{ in g mol}^{-1}}{(T/10^3 \text{ K})}\right)$$
$$\cdot \left[\int_{\in \text{envelope}} \frac{dP}{\rho} \bigg/ \left(\frac{G \cdot M_\oplus}{R_\oplus}\right)\right]$$
$$= 7.5 \cdot \left(\frac{\bar{\mu} \text{ in g mol}^{-1}}{(T/10^3 \text{ K})}\right) \cdot z; \quad (A7)$$

alternatively, this ratio expressed in common logarithm (log base 10) is

$$\log_{10}\left(\frac{P_{\text{env}}}{P_{\text{top}}}\right) = 3.3 \cdot \left(\frac{\bar{\mu} \text{ in g mol}^{-1}}{(T/10^3 \text{ K})}\right)$$
$$\cdot \left[\int_{\in \text{envelope}} \frac{dP}{\rho} \bigg/ \left(\frac{G \cdot M_\oplus}{R_\oplus}\right)\right]$$
$$= 3.3 \cdot \left(\frac{\bar{\mu} \text{ in g mol}^{-1}}{(T/10^3 \text{ K})}\right) \cdot z. \quad (A8)$$

We caution that the ideal gas approximation is generally good up to the $\sim 0.1$ GPa pressure level, above which the nonideal behavior, that is, significant intermolecular interactions, must be taken into account.

$P_{\text{top}}$ represents the edge of the exoplanet as determined by the mode of transit observation, where it is generally considered to be on the order of $\sim 10$ mbar or $\sim 10^3$ Pa equivalently. Then, substituting in Equation (A5) and using $\bar{\mu}_{H_2-He} \approx 2.3$ for an ideal gas envelope with uniform metallicity $Z$ at temperature $T$, we have

$$\log_{10}\left(\frac{P_{\text{env}}}{10^3 \text{ Pa}}\right) \approx \left(\frac{7.5/(1-Z)}{(T/10^3 \text{ K})}\right)$$
$$\cdot \left[\int_{\in \text{envelope}} \frac{dP}{\rho} \bigg/ \left(\frac{G \cdot M_\oplus}{R_\oplus}\right)\right]$$
$$= \left(\frac{7.5/(1-Z)}{(T/10^3 \text{ K})}\right) \cdot z. \quad (A9)$$

### A.2. Power-law EOS

$$P_{\text{power-law}} = \lambda \cdot \left(\frac{\rho}{1 \text{ g cc}^{-1}}\right)^{\Gamma}, \quad (A10)$$

where $\lambda$ has units of pressure and $\Gamma$ is a dimensionless power index. If $\lambda$ is constant,

$$\begin{aligned}
\int_{\in \text{envelope}} \frac{dP}{\rho} &= \int_{\in \text{envelope}} \lambda \cdot \Gamma \cdot \rho^{\Gamma-2} \cdot d\rho \\
&= \lambda \cdot \frac{\Gamma}{\Gamma - 1} \cdot (\rho_{\text{bottom}}^{\Gamma-1} - \rho_{\text{top}}^{\Gamma-1}) \\
&= \left(\frac{\Gamma}{\Gamma - 1}\right) \cdot \left(\frac{P_{\text{bottom}}}{\rho_{\text{bottom}}} - \frac{P_{\text{top}}}{\rho_{\text{top}}}\right). \quad (A11)
\end{aligned}$$

Typically, $\Gamma > 1$ and $\rho_{\text{bottom}} \gg \rho_{\text{top}}$. Thus, approximately,

$$\begin{aligned}
\int_{\in \text{envelope}} \frac{dP}{\rho} &\approx \lambda \cdot \left(\frac{\Gamma}{\Gamma - 1}\right) \cdot \rho_{\text{bottom}}^{\Gamma-1} \\
&= \left(\frac{\Gamma}{\Gamma - 1}\right) \cdot \left(\frac{P_{\text{bottom}}}{\rho_{\text{bottom}}}\right). \quad (A12)
\end{aligned}$$

Through manipulation, we obtain

$$\left(\frac{P_{\text{bottom}}}{\rho_{\text{bottom}}}\right) \approx \left(\frac{\Gamma - 1}{\Gamma}\right) \cdot \left[\int_{\in \text{envelope}} \frac{dP}{\rho} \bigg/ \left(\frac{G \cdot M_{\oplus}}{R_{\oplus}}\right)\right]$$
$$\cdot \left(\frac{G \cdot M_{\oplus}}{R_{\oplus}}\right) = \left(\frac{\Gamma - 1}{\Gamma}\right) \cdot z \cdot \left(\frac{G \cdot M_{\oplus}}{R_{\oplus}}\right), \quad (A13)$$

or equivalently,

$$\left(\frac{P_{\text{bottom}}}{\rho_{\text{bottom}}}\right) \approx \left(\frac{\Gamma - 1}{\Gamma}\right) \cdot \left[\int_{\in \text{envelope}} \frac{dP}{\rho} \bigg/ \left(\frac{G \cdot M_{\oplus}}{R_{\oplus}}\right)\right]$$
$$\cdot \left(\frac{62.56 \text{ GPa}}{1 \text{ g cm}^{-3}}\right) = \left(\frac{\Gamma - 1}{\Gamma}\right) \cdot z \cdot \left(\frac{62.56 \text{ GPa}}{1 \text{ g cm}^{-3}}\right). \quad (A14)$$

For isentropic metallic fluid hydrogen under high pressure, $\lambda \approx 200$ GPa and $\Gamma \approx 2$ (Becker et al. 2013, 2014, 2018; Militzer & Hubbard 2013; Weppner et al. 2015; Chabrier et al. 2019).[8]

In this particular case of $\Gamma = 2$ and $\lambda = 200$ GPa,

$$\left(\frac{\rho_{\text{bottom}}}{1 \text{ g cm}^{-3}}\right) \approx \left(\frac{2 - 1}{2}\right) \cdot \left[\int_{\in \text{envelope}} \frac{dP}{\rho} \bigg/ \left(\frac{G \cdot M_{\oplus}}{R_{\oplus}}\right)\right]$$
$$\cdot \left(\frac{62.56 \text{ GPa}}{200 \text{ GPa}}\right) = \left(\frac{2 - 1}{2}\right) \cdot z \cdot \left(\frac{62.56 \text{ GPa}}{200 \text{ GPa}}\right), \quad (A15)$$

then

$$\left(\frac{\rho_{\text{bottom}}}{1 \text{ g cm}^{-3}}\right) \approx (0.1564)$$
$$\cdot \left[\int_{\in \text{envelope}} \frac{dP}{\rho} \bigg/ \left(\frac{G \cdot M_{\oplus}}{R_{\oplus}}\right)\right] = (0.1564) \cdot z, \quad (A16)$$

then

$$P_{\text{env}} \approx (5 \text{ GPa})$$
$$\cdot \left[\int_{\in \text{envelope}} \frac{dP}{\rho} \bigg/ \left(\frac{G \cdot M_{\oplus}}{R_{\oplus}}\right)\right]^2 = (5 \text{ GPa}) \cdot z^2. \quad (A17)$$

We caution that this power-law approximation for isentropic metallic fluid hydrogen is generally good for $P_{\text{env}}$ greater than $\sim 1$ GPa pressure level, above which the hydrogen becomes significantly compressed and degenerate.

Taking into account the metallicity of the envelope, for a uniform admixture of mass fraction $Z = \sum Z_i$ of all heavier species $Z_i$ combined except hydrogen and helium in the isentropic metallic fluid hydrogen envelope, again let us assume volume additivity[8]:

$$\frac{1}{\bar{\rho}(P)} \approx \frac{1 - Z}{\sqrt{P/\lambda}} + \frac{Z}{\bar{\rho}_{\text{heavy}}}. \quad (A18)$$

If $Z$ is not too large and the heavier species are much denser than hydrogen–helium, then the second term in Equation (A18) can be dropped, and we arrive at the following approximate equation:

$$\frac{1}{\bar{\rho}(P)} \approx \frac{1 - Z}{\sqrt{P/\lambda}}, \quad (A19)$$

then, effectively,

$$P_{\text{mix}} \approx \lambda \cdot (1 - Z)^2 \cdot \left(\frac{\bar{\rho}}{1 \text{ g cc}^{-1}}\right)^2, \quad (A20)$$

and then the bottom pressure of the envelope becomes

$$P_{\text{env}} \approx \left(\frac{5 \text{ GPa}}{(1 - Z)^2}\right)$$
$$\cdot \left[\int_{\in \text{envelope}} \frac{dP}{\rho} \bigg/ \left(\frac{G \cdot M_{\oplus}}{R_{\oplus}}\right)\right]^2 = \left(\frac{5 \text{ GPa}}{(1 - Z)^2}\right) \cdot z^2. \quad (A21)$$

For back-of-envelope calculation of $P_{\text{env}}$, we could use the following criterion to quickly estimate its value: always take the smaller value of Equations (A8) and (A17) (or, equivalently, the smaller value of Equations (A9) and (A21)).

## Appendix B
## An Approximate Analytic EOS for Ices/Fluids

We define a dimensionless density $\eta$ (also known as the compression factor). It is the ratio of density at any condition versus ice density at ambient conditions ($\rho_0 \approx 1$ g cm$^{-3}$):

$$\eta \equiv \rho/\rho_0 = \rho/(1 \text{ g cm}^{-3}). \quad (B1)$$

We define a dimensionless temperature $t$ as temperature normalized by the critical temperature $T_c$ (647.1 K for $H_2O$):

$$t \equiv T/T_c = T/(647.096 \text{ K}). \quad (B2)$$

Then, the pressure (in GPa) is approximately (Tsien 1965)

$$\frac{P}{\text{GPa}} \approx \eta^4 - \eta^2 + t \cdot \eta \cdot \exp\left(\frac{\eta^2 - 1}{\eta^2 + 1}\right). \quad (B3)$$

Equation (B3) captures the increase of pressure due to the electronic Coulomb repulsion, the critical behavior, and the thermal effects (including both ideal gas behavior under low-density, high-temperature conditions and the atomic vibrations in crystal lattices under high-density conditions). The temperature dependence term in Equation (B3) works for the following reasons:

1. At low density: according to ideal gas law, $P = n \cdot k_B \cdot T$, where $n$ is the number density of molecules, where each $H_2O$ molecule contains three atoms.

2. At high density: $P_{\text{thermal}} \approx 3 \cdot \tilde{n} \cdot k_B \cdot T$, where $\tilde{n}$ is the number density of atoms instead. The factor of three comes in here because each atom can vibrate in three dimensions, $x$, $y$, $z$, and according to equipartition of energy, each vibration mode possesses the same amount of energy.

The temperature dependence term in Equation (B3) can be tuned to higher accuracy if needed, by inserting one or two extra parameters into the exponential term. However, the temperature dependence term in Equation (B3) is usually small compared to the first two terms for condensed phases ($\gtrsim 1$ g cm$^{-3}$). Thus, for more simplicity, one can even drop the temperature dependence term completely and arrive at an even simpler analytic EOS:

$$\frac{P}{\text{GPa}} \approx \eta^4 - \eta^2. \tag{B4}$$

Equation (B4) can approximate the behavior of $H_2O$, $NH_3$, and $CH_4$, as well as their mixtures of any proportion, from $\sim 1$ up to $\sim 5$ g cm$^{-3}$, which covers the mass–radius regime of interest of any water worlds below $\sim 20\ M_\oplus$.

## Appendix C
## Exact Formulation of Water EOS

### C.1. Critical Point

Figure 11 shows the $H_2O$ EOS of 200–1000 K and 0–2.5 g cm$^{-3}$. Like many other substances, $H_2O$ has a C.P. Below the C.P., the liquid phase and gas phase separate and are distinguishable from each other with a finite difference in their densities or volumes. Above the C.P., the transition between the liquid-like high-density region and gas-like low-density region becomes continuous, and the fluid there is considered supercritical. The C.P. for $H_2O$ is quoted as follows (IAPWS 2011, 2018, 2019):

$$\rho_c = 0.322 \text{ g cm}^{-3}$$
$$T_c = 647.096 \text{ K} = 373.946°C$$
$$P_c = 2.2064 \cdot 10^7 \text{ Pa} = 220.64 \text{ bar}$$
$$n_c = 0.322/18 = 0.018 \text{ mol cm}^{-3}.$$

Here $n_c$ stands for the molar density (mol cm$^{-3}$) of water at its C.P.

### C.2. EOS of Ices

To the left of the "wedge"-shaped liquid region, at low temperatures, various ice phases exist with approximately quantized densities, that is, each possessing an almost unique density with weak dependence on pressure/temperature, which is omitted from our graphs. These ice phases, including Ice II, Ice III, Ice V, and Ice VI, can be considered transitional or intermediate phases, in between the ordinary Ice Ih and Ice VII —the predominating high-pressure ice phase from $\sim 2$ to $\sim 40$ GPa.

This quantization of densities can be compared to the quantization of energy levels in a cold quantum system such as a harmonic oscillator. Since at relatively low temperature, there are only certain well-defined spatial configurations and orientations of molecules in the crystal lattices that could minimize the energy of the whole system. This quantization is in contrast to the continuous density changes allowed in the liquid phase.

These intermediate ice phases are only slightly denser than their corresponding liquid in equilibrium, except Ice Ih, which is less dense. Thus, if cold enough (below 273 K), the water worlds, such as Pluto and Ganymede, are always covered with an Ice Ih crust at the very top. Then, deeper within the interior, there could exist liquid ocean of various depths, if there is a heat source in the interior to maintain its temperature above freezing, and/or thin layers of Ice (II, III, V) + Ice VI. Then, eventually, above a few GPa, Ice VII is found.

To summarize the behavior of these ice phases:

1. Ice Ih exists from 0 up to 0.2 GPa.
2. Ice (II, III, V) have similar densities, and they exist from 0.2 up to 0.6 GPa. In approximation, these three can be treated as one.
3. Ice VI exists from 0.6 up to 2 GPa. Ice VII exists from 2 GPa onward.

The Ice VI–Ice VII–liquid triple point (*t*2) is to be found at

$$T_{t2} = 355 \text{ K}$$
$$P_{t2} = 2.216 \text{ GPa} = 2.216 \cdot 10^4 \text{ bar}$$
$$\rho_{\text{VI}} = 1.31 \text{ g cm}^{-3} \text{ for Ice VI}$$
$$\rho_{\text{VII}} = 1.567 \text{ g cm}^{-3} \text{ for Ice VII}$$
$$\rho_{\text{Liq}} = 1.35 \text{ g cm}^{-3} \text{ for liquid}$$
$$n_{\text{VI}} \approx 1.31/18 \approx 0.0728 \text{ mol cm}^{-3} \text{ for Ice VI}$$
$$n_{\text{VII}} \approx 1.567/18 \approx 0.0871 \text{ mol cm}^{-3} \text{ for Ice VII}$$
$$n_{\text{Liq}} \approx 1.35/18 \approx 0.075 \text{ mol cm}^{-3} \text{ for liquid,}$$

where $n$ is the molar density of each phase in (mol cm$^{-3}$) at this triple point.

According to Bezacier et al. (2014), the EOS parameters for Ice VI and Ice VII are as follows. Note that the subscript "0" suggests extrapolation of the parameter under ambient conditions ($P_0 = 1$ bar and $T_0 = 300$ K). These parameters are obtained by fit to the Birch–Murnagham second-order (BM2) EOS (Birch 1947, 1952) with temperature-dependent corrections as follows:

Density correction:

$$\rho(P, T) = \rho_0(P, T_0) \cdot \exp(-\alpha_0 \cdot (T - T_0)). \tag{C1}$$

Bulk modulus correction:

$$K(T) = K_0 + (T - T_0) \cdot \left(\frac{\partial K}{\partial T}\right)_P. \tag{C2}$$

The melting curve of both Ice VI and Ice VII is fit to a Simon–Glatzel form of the equation (Simon & Glatzel 1929):

$$P_M = P_0 + a \cdot \left(\left(\frac{T_M}{T_0}\right)^c - 1\right), \tag{C3}$$

where $a$ and $c$ are constants to be determined. More recent development on the parameterization of the melting curve involves the Kechin equation (Stishov 1975; Kechin 1995).

### C.2.1. Ice VI

Thermodynamic parameters of Ice VI (Bezacier et al. 2014):

$$\rho_0 \approx 1.27 \text{ g cm}^{-3}$$
$$K_0 \approx 14 \text{ GPa}$$
$$\alpha_0 \approx 14.6 \cdot 10^{-5} \text{ K}^{-1}$$
$$c_v \approx 2.5 \frac{\text{J K}^{-1}}{g}$$
$$\gamma = \frac{\alpha \cdot K_T}{c_v \cdot \rho} \approx 0.65. \quad (C4)$$

Melting curve of Ice VI (IAPWS 2011):

$$P_M \approx 0.63 + 0.68 \cdot \left(\left(\frac{T_M}{273 \text{ K}}\right)^{4.6} - 1\right) \text{GPa}. \quad (C5)$$

### C.2.2. Ice VII

Thermodynamic parameters of Ice VII (Bezacier et al. 2014):

$$\rho_0 \approx 1.45 \text{ g cm}^{-3}$$
$$K_0 \approx 20 \text{ GPa}$$
$$\alpha_0 \approx 11.6 \cdot 10^{-5} \text{ K}^{-1}$$
$$c_v \approx 2 \frac{\text{J K}^{-1}}{g}$$
$$\gamma = \frac{\alpha \cdot K_T}{c_v \cdot \rho} \approx 0.81. \quad (C6)$$

Melting curve of Ice VII (Frank et al. 2004), fit to experimental data up to 60 GPa:

$$P_M \approx 2.17 + 0.764 \cdot \left(\left(\frac{T_M}{355 \text{K}}\right)^{4.32} - 1\right) \text{GPa}. \quad (C7)$$

The density of Ice VII along the melting curve is (Frank et al. 2004)

$$\rho_M \approx 1.45 + 0.4 \cdot \left(1 - \exp\left(-0.0743 \frac{P_M}{\text{GPa}}\right)\right)$$
$$+ 2.8 \cdot \left(1 - \exp\left(-0.0061 \frac{P_M}{\text{GPa}}\right)\right) \text{g cm}^{-3}. \quad (C8)$$

Thus, a good rule of thumb for Grüneisen parameter $\gamma$ in both Ice VI and Ice VII is that $\gamma \approx 0.7$–$0.8$.

### C.3. EOS of the Superionic State

This subsection deals with calculating the thermodynamic parameters of superionic ice, from the standpoint of the Birch–Murnagham second-order (BM2) EOS (Birch 1947, 1952) for cold compression:

$$P_{\text{BM2}}(\rho) = \frac{3}{2} \cdot K_0 \cdot \left(\left(\frac{\rho}{\rho_0}\right)^{7/3} - \left(\frac{\rho}{\rho_0}\right)^{5/3}\right). \quad (C9)$$

1. $\rho_0$ is the reference density;
2. $K_0$ is the reference isothermal bulk modulus.

The best fit of Equation (C9) to the experimental and ab initio simulation data of superionic ice (Millot et al. 2018) gives the following values:

$$K_0 = 110 \text{ GPa}$$
$$\rho_0 = 2.1 \text{ g cm}^{-3}.$$

The superionic ice is also called Ice XVIII (Chaplin 2019). It is now experimentally verified (Millot et al. 2018, 2019). The fluid–Ice VII–Ice XVIII triple point (t3) is important for plotting the $H_2O$ phase diagram beyond a temperature of ~1000 K and density of ~2.5 g cm$^{-3}$ or pressure of ~40 GPa (see Figure 12):

$$\rho_{t3} \sim 2.5 \text{ g cm}^{-3} \text{ for fluid}$$
$$T_{t3} \sim 1000 \text{ K}$$
$$P_{t3} \sim 40 \text{ GPa} \sim 4 \cdot 10^5 \text{ bar}$$
$$n_{t3} \sim 2.5/18 \sim 0.13 \text{ mol cm}^{-3} \text{ for fluid}.$$

Calculations show that ammonia ($NH_3$) and its mixture with water ($H_2O$) also possess a superionic phase (Cavazzoni et al. 1999; Bethkenhagen et al. 2013, 2015; Jiang et al. 2017) at high pressure and temperature.

### C.4. EOS of Fluids

The International Association for the Properties of Water and Steam (IAPWS; IAPWS 2011, 2018) compiles and publishes analytic EOSs of fluid $H_2O$, which includes physical and chemical properties of liquid, gas, and supercritical fluid, based on new experimental data each year. Our calculation of fluid is based on the IAPWS formulation.

The IAPWS formulation is based on the principle of corresponding states. That is, the physical quantities such as temperature, density, and pressure are normalized and non-dimensionalized by its value at exactly the C.P. Following from this principle, an EOS can be expressed as

$$f\left(\frac{\rho}{\rho_c}, \frac{T_c}{T}, \frac{P}{P_c}\right) = 0, \quad (C10)$$

where $\rho_c$ is the density at the C.P., $T_c$ is the temperature at the C.P., and $P_c$ is the pressure at the C.P. As a general remark, J. van der Waals's idea of corresponding states was anticipated by Mendeleev, who pointed to the usefulness of comparing volumes not at the boiling points but at temperatures when the cohesion of the liquid molecules is close to zero (Karapetyants 1978). Moreover, Equation (C10) can be used for similar species if the intermolecular interactions among these species are similar.

Anchored at the C.P., the IAPWS fluid EOS formulation is then built on the power series expansion from the ideal gas EOS. The following is a brief summary of the IAPWS formulation (IAPWS 2011, 2018). First of all, we define two independent variables: reduced density $\delta$ (dimensionless),

$$\delta = \frac{\rho}{\rho_c}, \quad (C11)$$

and inverse reduced temperature $\tau$ (dimensionless),

$$\tau = \frac{T_c}{T}. \quad (C12)$$

All other thermodynamic variables are expressed as functions in them ($\delta$ and $\tau$). Then, we write the specific Helmholtz free energy $f$ as

$$f = u - T \cdot s. \quad (C13)$$

All other thermodynamic quantities can be derived from $f$ through appropriate partial derivatives. Equation (C13) can be cast into a dimensionless form as

$$\psi = \frac{f}{R \cdot T} = \psi^o(\delta, \tau) + \psi^r(\delta, \tau), \quad (C14)$$

where $\psi^o$ is the ideal gas part and $\psi^r$ is the residual part.

$\psi^o$ and $\psi^r$ are each expressed as an analytic formula in terms of $\delta$ and $\tau$, with coefficients determined by best fit to experimental data:

$$\psi^o[\delta, \tau] = \ln \delta + n_1^o + n_2^o \cdot \tau + n_3^o \cdot \ln \tau \\ + \sum_{i=4}^{8} n_i^o \cdot \ln[1 - \exp(-\gamma_i^o \cdot \tau)], \quad (C15)$$

where $n_i^o$ and $\gamma_i^o$ are dimensionless coefficients. Likewise,

$$\psi^r[\delta, \tau] = \sum_{i=1}^{7} n_i \cdot \delta^{d_i} \cdot \tau^{t_i} + \sum_{i=8}^{51} n_i \cdot \delta^{d_i} \cdot \tau^{t_i} \cdot \exp(-\delta^{c_i}) \\ + \sum_{i=52}^{54} n_i \cdot \delta^{d_i} \cdot \tau^{t_i} \cdot \exp(-\alpha_i \cdot (\delta - \epsilon_i)^2 - \beta_i \cdot (\tau - \gamma_i)^2) \\ + \sum_{i=55}^{56} n_i \cdot \Delta^{b_i} \cdot \delta \cdot \psi, \quad (C16)$$

with various coefficients introduced to fine-tune the formula to fit the experimental data:

$$\Delta = \theta^2 + B_i \cdot [(\delta - 1)^2]^{\alpha_i} \\ \theta = (1 - \tau) + A_i \cdot [(\delta - 1)^2]^{\frac{1}{2 \cdot \beta_i}} \\ \psi = \exp(-C_i \cdot (\delta - 1)^2 - D_i \cdot (\tau - 1)^2), \quad (C17)$$

where the specific gas constant for this mass-based formulation is taken as

$$R = 0.46151805 \text{ kJ} \cdot \text{kg}^{-1} \cdot \text{K}^{-1}. \quad (C18)$$

Much of the complexity is introduced to describe the behavior of $H_2O$ around its C.P., where the pressure changes tremendously over a very small density and temperature range. Beyond C.P., the distinction between the liquid and vapor phases vanishes, and water is supercritical, existing as small but liquid-like hydrogen-bonded clusters dispersed within a gas-like phase (Chaplin 2019).

The pressure $P$ can be calculated from the specific Helmholtz free energy $f$ through the fundamental thermodynamic relation

$$P = \rho^2 \cdot \left(\frac{\partial f}{\partial \rho}\right)_T \quad (C19)$$

and be expressed in dimensionless variables as

$$\frac{P(\delta, \tau)}{\rho \cdot R \cdot T} = 1 + \delta \cdot \left(\frac{\partial \psi^r}{\partial \delta}\right)_\tau. \quad (C20)$$

Likewise, all other thermodynamic functions and variables can be derived from the specific Helmholtz free energy $f$ through appropriate partial derivatives.

Internal energy (specific) $u$:

$$\frac{u(\delta, \tau)}{R \cdot T} = \tau \cdot \left(\left(\frac{\partial \psi^o}{\partial \tau}\right)_\delta + \left(\frac{\partial \psi^r}{\partial \tau}\right)_\delta\right). \quad (C21)$$

Entropy (specific) $s$:

$$\frac{s(\delta, \tau)}{R} = \tau \cdot \left(\left(\frac{\partial \psi^o}{\partial \tau}\right)_\delta + \left(\frac{\partial \psi^r}{\partial \tau}\right)_\delta\right) - \psi^o - \psi^r. \quad (C22)$$

Enthalpy (specific) $h$:

$$\frac{h(\delta, \tau)}{R \cdot T} = 1 + \tau \cdot \left(\left(\frac{\partial \psi^o}{\partial \tau}\right)_\delta + \left(\frac{\partial \psi^r}{\partial \tau}\right)_\delta\right) + \delta \cdot \left(\frac{\partial \psi^r}{\partial \delta}\right)_\tau. \quad (C23)$$

The specific internal energy $u$ and specific entropy $s$ of saturated liquid at the gas–liquid–Ice Ih triple point are set to zero to determine the constants of integration. The gas–liquid–Ice Ih triple point ($t1$) is (Linstrom & Mallard 2020)

$$\rho_{t1} = 0.99979 \approx 1 \text{ g cm}^{-3} \text{ for liquid} \\ T_{t1} = 273.16 \text{ K} = 0.01°\text{C} \\ P_{t1} = 611.655 \text{ Pa} \approx 0.006 \text{ bar} \\ n_{t1} \approx 1/18 \approx 0.0555 \text{ mol cm}^{-3} \text{ for liquid.}$$

Because the IAPWS formulation is based on temperature $T$ and density $\rho$ as independent variables, that is why the temperature–density graph is first constructed (Figures 11 and 12), in order to illustrate the $H_2O$ EOS.

Lastly, the applicable range of the IAPWS formulation is the entire stable fluid regime that includes liquid, vapor, and supercritical fluid beyond the C.P. It is experimentally verified for temperature $T$ in between 273 and 1273 K and pressures $P$ up to about $\sim$1 GPa. However, tests and comparison with experiments have shown that this IAPWS analytic formulation can be safely extrapolated for at least density and enthalpy of undissociated $H_2O$ up to $\sim$5000 K and $\sim$100 GPa (IAPWS 2018). The next subsection concerns the dissociation/ionization of fluid $H_2O$.

### C.5. IAPWS Ionization Constant of $H_2O$ Fluids

The coloring of the fluids shown in Figures 11 and 12 represents the ionization constant ($pK_w$) of $H_2O$, according to the analytic formula provided by IAPWS (2019). By definition, $pK_w$ is the water autoionization constant[12]:

$$K_w = [H_3O+][OH-] \\ pK_w = -\log[K_w] \\ pK_w = \text{pH} + \text{pOH} \\ \text{pH} = -\log[H_3O+] \\ \text{pOH} = -\log[OH-]. \quad (C24)$$

Both $[H_3O+]$ and $[OH-]$ are measured in mol l$^{-1}$ (molarity). As commonly known, $pK_w = 14$ is for pure $H_2O$ under ambient conditions. That is why pH = 7 and pOH = 7 represent the neutral pure water under ambient conditions, neither acidic nor basic. However, $pK_w$ will change under other temperature and density/pressure conditions. When $K_w$ crosses over $\sim$1 (unity), or equivalently, when $pK_w$ crosses below 0, then the $H_2O$ fluid becomes significantly ionized, and thus it should be considered ionic fluid from there on. It is relevant when considering chemical reactions/equilibrium under those temperature and density/pressure conditions other than ambient conditions.

---

[12] https://www.chemteam.info/AcidBase/Kw.html

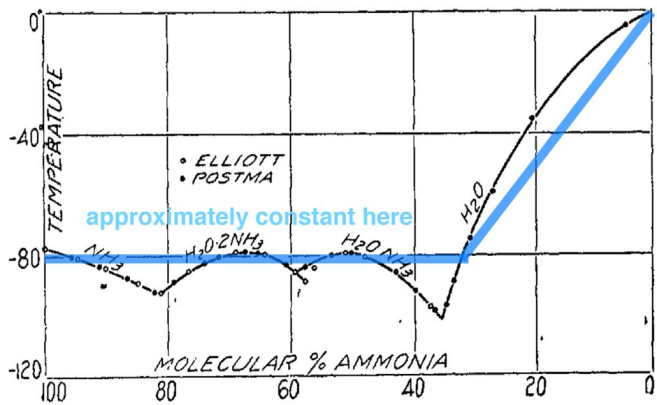

**Figure 15.** Freezing point curve of water–ammonia mixture (modified from Elliott 1924).

### C.6. Water–Ammonia Mixture

Let us consider the details for $H_2O$–$NH_3$ mixture in Figure 11. We do a simple approximation of the freezing point curve of the $H_2O$–$NH_3$ mixture: linear drop from 0°C to −80°C for zero up to ∼one-third (∼1/3 by weight) ammonia in the mixture, and then staying roughly constant (−80°C) from there on when higher percentages of ammonia are added to the mixture (Elliott 1924; IAPWS 2001; see Figure 15).

## Appendix D
## Mathematica SourceCode

```
(*Set the working directory to notebook directory*)

SetDirectory [NotebookDirectory [ ]] ;

(*Start of the formal manipulate code:*)

Manipulate[
 plt = Grid[{
  {plt3 = Dynamic@Histogram[{
   (# [[27]] *317.8) & /@

   Select [data0 [[2 ;;]] ,
    mmin ⇐ (# [[27]] *317.8) ⇐ mmax && NumberQ [# [[28]] ]  &&
     NumberQ [# [[29]] ]  && rmin ⇐ (# [[30]] *11.2) ⇐ rmax &&
     NumberQ [# [[31]] ]  &&
     NumberQ [# [[32]] ]  && # [[31]] /# [[30]] ⇐ / [Sigma] Rpercent/
     100. && # [[32]] /# [[30]] ⇐ / [Sigma] Rpercent/
     100. && # [[28]] /# [[27]] ⇐ / [Sigma] Mpercent/
     100. && # [[29]] /# [[27]] ⇐ / [Sigma] Mpercent/
     100. && # [[30]] ^4/# [[27]]  > 0.01 &&
     Porb [[1]] ⇐ # [[20]] ⇐ Porb [[2]] &&
     Teq [[1]] ⇐ # [[39]] ⇐ Teq [[2]] &&
     Teff [[1]] ⇐ # [[2]] ⇐ Teff [[2]] &&
     eccentricity [[1]] ⇐ # [[21]] ⇐ eccentricity [[2]] &&
     mstar [[1]] ⇐ # [[8]] ⇐ mstar [[2]] &&
     rstar [[1]] ⇐ # [[11]] ⇐ rstar [[2]] &&
     aorb [[1]] ⇐ # [[24]] ⇐ aorb [[2]] &&
     First[FeHdex] ⇐ # [[5]] ⇐ Last[FeHdex]  &],
 If[
  add1, # [[2]] & /@
```

(Continued)

```
   Select [np2,
    If [filter1,
     mmin ⇐ (# [[2]] ) ⇐ mmax && rmin ⇐ (# [[5]] ) ⇐ rmax &&
      Teq [[1]] ⇐ (# [[8]] ) ⇐ Teq [[2]]  &
     mmin ⇐ (# [[2]] ) ⇐ mmax &&
      rmin ⇐ (# [[5]] ) ⇐ rmax &]], {}]
 },
 (*{0.1},*)
 If [xscale,
 {"Log", {10^
   Range[Log[10, mmin],
    Log[10, mmax], (Log[10, mmax] - Log[10, mmin])/
    histmassbin]}}, {mmin, mmax, (mmax—mmin)/histmassbin}],
  AspectRatio → 1/5, BarOrigin → Bottom,
  ChartLayout → "Stacked",
  PlotRangePadding → None,
  ImageSize → {450, Automatic},
  (*ImagePadding /[Rule]{{left,right},{bottom,top}}*)

  ImagePadding → {{50, 0}, {0, 50}},
  Frame → True,
  (*FrameLabel → {{left,right},{bottom,top}}*)

  FrameTicks → All,
  FrameLabel → {{"counts", ""}, {"",
   "Histogram of (/!/(/* SubscriptBox[/(M/), /
/(p/)]/)//!/(/* SubscriptBox[/(M/), /(/[Earth]/)]/))"}},
  FrameStyle → Directive [14, Gray, Thicknes s[0.005]] ],

 plt4 = Dynamic@Histogram[{
  (# [[30]] *11.2)/(# [[27]] *317.8)^(1/4) & /@

  Select [data0 [[2 ;;]],
   mmin ⇐ (# [[27]] *317.8) ⇐ mmax && NumberQ [# [[28]] ] &&
    NumberQ [# [[29]] ]  && rmin ⇐ (# [[30]] *11.2) ⇐ rmax &&
    NumberQ [# [[31]] ]  &&
    NumberQ [# [[32]] ]  && # [[31]] /# [[30]] ⇐ / [Sigma] Rpercent/
    100. && # [[32]] /# [[30]] ⇐ / [Sigma] Rpercent/
    100. && # [[28]] /# [[27]] ⇐ / [Sigma] Mpercent/
    100. && # [[29]] /# [[27]] ⇐ / [Sigma] Mpercent/
    100. && # [[30]] ^4/# [[27]]  > 0.01 &&
    Porb [[1]] ⇐ # [[20]] ⇐ Porb [[2]] &&
    Teq [[1]] ⇐ # [[39]] ⇐ Teq [[2]] &&
    Teff [[1]] ⇐ # [[2]] ⇐ Teff [[2]] &&
    eccentricity [[1]] ⇐ # [[21]] ⇐ eccentricity [[2]] &&
    mstar [[1]] ⇐ # [[8]] ⇐ mstar [[2]] &&
    rstar [[1]] ⇐ # [[11]] ⇐ rstar [[2]] &&
    aorb [[1]] ⇐ # [[24]] ⇐ aorb [[2]] &&
    First[FeHdex] ⇐ # [[5]] ⇐ Last[FeHdex]  &],
 If[
 add1, (# [[5]] )/(# [[2]] )^(1/4) & /@

  Select [np2,
   If [filter1,
    mmin ⇐ (# [[2]] ) ⇐ mmax && rmin ⇐ (# [[5]] ) ⇐ rmax &&
     Teq [[1]] ⇐ (# [[8]] ) ⇐ Teq [[2]]  &
    mmin ⇐ (# [[2]] ) ⇐ mmax &&
     rmin ⇐ (# [[5]] ) ⇐ rmax &]], {}]
 },
 {1/histzetabin},
 AspectRatio → 1/GoldenRatio,
 (*ScalingFunctions /[Rule]{"Log","Log"},*)

 ChartLayout → "Stacked",
 ImageSize → {Automatic, 150},
 (*ImagePadding /[Rule]{{left,right},{bottom,top}}*)
```

(Continued)

```
    ImagePadding → {{0, 50}, {0, 70}},
    Frame → True,
    (*FrameLabel → {{left,right},{bottom,top}}*)

    FrameLabel → {{"", "counts"}, {"",
      "Histogram of /[Zeta] = /!/(/*FractionBox[/
/(R/), /(p/)]//*SubscriptBox[/
/(R/), /(/[Earth]/)]/)/), SuperscriptBox[/
/((/*SubscriptBox[/(M/), /(P/)]//*SubscriptBox[/(M/),/
/(/[Earth]/)]/)/), /(1/4/)]] /)"}},
    FrameTicks → All,
    FrameStyle → Directive [9, Gray, Bold, Thicknes s[0.007]] ] }
,
  {plt1 = Show[
    If[env1,
    Which[
    env2 == 1,
    {(*Density Plot for Fe-Silicates Contour Mesh,
    approximated by Power-Series in lg [mass] *)

    DensityPlot [
      (r −(10. ˆ(−0.11408792224566819` + 0.27851883673695` x−
        0.01997874049680844` x^2-0.002490304269884624` x^3 +
        0.00007525048500183394` x^4-0.00007162041164677924` x^5-
0.00003393158521958243` x^6 +
        8.599995554646332`*ˆ−7 x^7 +
        1.132375249329131`*^-6 x^8 +
        2.2299345660512832`*ˆ−8 x^9-1.0475165171649914`*ˆ−8 x^10) /.
      x → Log10[
        m]))/((10. ˆ(0.020013868549526272` +
        0.29811170324848235` x−
        0.02012734730157388` x^2-0.0052918215948260265` x^3-
0.0003311775031243655` x^4 +
        0.00004856681718363753` x^5-0.00001245509278944841` x^6-
1.3074832660503483`*ˆ-6 x^7 +
        8.211419885278952`*ˆ−7 x^8 +
        3.47368749025812`*ˆ−8 x^9-1.1251826465596989`*ˆ−8 x^10) /.
      x → Log10[
        m]) - (10. ˆ(-0.11408792224566819` +
        0.27851883673695` x−−0.01997874049680844` x^2-
0.002490304269884624` x^3 +
        0.00007525048500183394` x^4−0.00007162041164677924` x^5-
0.00003393158521958243` x^6 +
        8.599995554646332`*ˆ−7 x^7 +
        1.132375249329131`*ˆ−6 x^8 +
        2.2299345660512832`*ˆ−8 x^9−1.0475165171649914`*ˆ−8 x^10) /.
    x → Log10[m])),
      {m, mmin, mmax}, {r, rmin, rmax},

    ScalingFunctions → {If [xscale,"Log", None],
      If[yscale, "Log", None]},
    MeshFunctions → {#3 &}, Mesh → {{0.2, 0.4, 0.6, 0.8}},
    PlotPoints → ControlActive[10, 30],

    RegionFunction →
    Function[{m, r,
      z}, (10. ˆ(−0.11408792224566819` + 0.27851883673695` x−−
        0.01997874049680844` x^2-0.002490304269884624` x^3 +
        0.00007525048500183394` x^4-0.00007162041164677924`
x^5−0.00003393158521958243` x^6 +
        8.599995554646332`*ˆ−7 x^7 +
        1.132375249329131`*ˆ−6 x^8 +
        2.2299345660512832`*ˆ−8 x^9-1.0475165171649914`*ˆ−8 x^10) /.
    x → Log10[m]) <
      r < (10. ˆ(0.020013868549526272` +
        0.29811170324848235` x−−
        0.02012734730157388` x^2-0.0052918215948260265` x^3-
0.0003311775031243655` x^4 +
        0.00004856681718363753` x^5-0.00001245509278944841` x^6-
1.3074832660503483`*ˆ-6 x^7 +
        8.211419885278952`*ˆ−7 x^8 +
        3.47368749025812`*ˆ−8 x^9-1.1251826465596989`*ˆ−8 x^10)/.
  x → Log10[m])],
    BoundaryStyle → None,
    ColorFunction → (Opacity[1., Blend[{Gray, White}, #]] &),
    ColorFunctionScaling → False,
    Exclusions → None],

    (*Density Plot for Silicates-H2O Contour Mesh,
    approximated by Power-Series in lg [mass] *)

    DensityPlot [(r - (10. ˆ(0.020013868549526272` +
      0.29811170324848235` x−
      0.02012734730157388` x^2-0.0052918215948260265` x^3-
0.0003311775031243655` x^4 +
      0.00004856681718363753` x^5−0.00001245509278944841`
x^6-1.3074832660503483`*ˆ-6 x^7 +
      8.211419885278952`*ˆ−7 x^8 +
      3.47368749025812`*ˆ−8 x^9−1.1251826465596989`*ˆ−8 x^10) /.
    x → Log10[
      m]))/((10. ˆ(0.13666292574887867` +
      0.27183702181443314` x−
      0.007134024332627119` x^2−0.0021407416433092126`
x^3−0.0022608931475693915` x^4-0.0002516518649610248` x^5 +
      0.00011968169122553435` x^6 +
      0.000011663496987412905` x^7−3.536434693875541`*ˆ-6
x^8-1.6848230313524644`*ˆ−7 x^9 +
      4.4044933682275176`*ˆ−8 x^10) /.
    x → Log10[
      m]) - (10. ˆ(0.020013868549526272` +
      0.29811170324848235` x−
      0.02012734730157388` x^2−0.0052918215948260265`
x^3−0.0003311775031243655` x^4 +
      0.00004856681718363753` x^5−0.00001245509278944841`
x^6−1.3074832660503483`*ˆ-6 x^7 +
      8.211419885278952`*ˆ−7 x^8 +
      3.47368749025812`*ˆ−8 x^9-1.1251826465596989`*ˆ−8 x^10) /.
x → Log10[m])),
    {m, mmin, mmax}, {r, rmin, rmax},

    ScalingFunctions → {If [xscale, "Log", None],
      If[yscale, "Log", None]},
    MeshFunctions → {#3 &}, Mesh → {{0.2, 0.4, 0.6, 0.8}},
    PlotPoints → ControlActive[10, 30],

    RegionFunction →
    Function[{m, r,
      z}, (10. ˆ(0.020013868549526272` +
      0.29811170324848235` x−
      0.02012734730157388` x^2-0.0052918215948260265` x^3-
0.0003311775031243655` x^4 +
      0.00004856681718363753` x^5-0.00001245509278944841` x^6-
1.3074832660503483`*ˆ-6 x^7 +
      8.211419885278952`*ˆ−7 x^8 +
      3.47368749025812`*ˆ−8 x^9-1.1251826465596989`*ˆ−8 x^10) /.
    x → Log10[m]) <
      r < (10. ˆ(0.13666292574887867` +
      0.27183702181443314` x−
      0.007134024332627119` x^2-0.0021407416433092126` x^3-
0.0022608931475693915` x^4-0.0002516518649610248` x^5 +
      0.00011968169122553435` x^6 +
```

(Continued)

```
    0.000011663496987412905` x^7-3.536434693875541`*^-6 x^8-
1.6848230313524644`*^−7 x^9 +
    4.4044933682275176`*^−8 x^10) /. x → Log10[m])],
  BoundaryStyle → None,
  ColorFunction → (Opacity[1., Blend[{Gray, White}, #]] &),
  ColorFunctionScaling → False,
  Exclusions → None],

(∗Density Plot for Envelope-H2O Contour Mesh,
  approximated by Power-Series in lg [mass] ∗)

  DensityPlot [(1/(10. ˄(0.13666292574887867` +
      0.27183702181443314` x—
    0.007134024332627119` x^2-0.0021407416433092126` x^3-
0.0022608931475693915` x^4-0.0002516518649610248` x^5+
      0.00011968169122553435` x^6 +
      0.000011663496987412905` x^7-3.536434693875541`*^-6 x^8-
1.6848230313524644`*^−7 x^9 +
      4.4044933682275176`*^−8 x^10) /. x → Log10[m]) -
    1/r)∗m,
  {m, mmin, mmax}, {r, rmin, rmax},

  ScalingFunctions → {If [xscale, "Log", None],
    If[yscale, "Log", None]},
  MeshFunctions → {#3 &},
  Mesh → {{0.1, 0.2, 0.3, 0.4, 0.5, 0.6, 0.7, 0.8, 0.9, 1.,
    1.5, 2., 2.5, 3.}},
  PlotPoints → ControlActive[10, 50],

  RegionFunction →
   Function[{m, r,
    z}, (10. ˄(0.13666292574887867` +
      0.27183702181443314` x—
    0.007134024332627119` x^2-0.0021407416433092126` x^3-
0.0022608931475693915` x^4-0.0002516518649610248` x^5+
      0.00011968169122553435` x^6 +
      0.000011663496987412905` x^7-3.536434693875541`*^-6 x^8-
1.6848230313524644`*^−7 x^9 +
      4.4044933682275176`*^−8 x^10) /. x → Log10[m]) <
    r],
  BoundaryStyle → None,
  ColorFunction → (Opacity[1., Blend[{Gray, White}, #]] &),
  ColorFunctionScaling → False,
  Exclusions → None]},

env2 == 2,
{(∗Density Plot for Fe-Silicates Contour Mesh,
  approximated by Power-Series in lg [mass] ∗)

DensityPlot [
  (r - (10. ˄(-0.11408792224566819` + 0.27851883673695` x—
    0.01997874049680844` x^2-0.002490304269884624` x^3 +
    0.00007525048500183394` x^4-0.00007162041164677924` x^5-
0.00003393158521958243` x^6 +
    8.589995554646332`*^−7 x^7 +
    1.132375249329131`*^-6 x^8 +
    2.2299345660512832`*^−8 x^9-1.0475165171649914`*^−8 x^10) /.
    x → Log10[
     m]))/((10. ˄(0.020013868549526272` +
      0.29811170324848235` x—
    0.02012734730157388` x^2-0.0052918215948260265` x^3-
0.0003311775031243655` x^4 +
    0.00004856681718363753` x^5-0.00001245509278944841` x^6-
1.3074832660503483`*^-6 x^7 +
```

```
    8.211419885278952`*^−7 x^8 +
    3.47368749025812`*^−8 x^9-1.1251826465596989`*^−8 x^10) /.
    x → Log10[
     m]) - (10. ˄(-0.11408792224566819` +
      0.27851883673695` x—0.01997874049680844` x^2-
0.002490304269884624` x^3 +
      0.00007525048500183394` x^4-0.00007162041164677924` x^5-
0.00003393158521958243` x^6 +
      8.589995554646332`*^−7 x^7 +
      1.132375249329131`*^-6 x^8 +
      2.2299345660512832`*^−8 x^9-1.0475165171649914`*^−8 x^10) /.
x → Log10[m])),
  {m, mmin, mmax}, {r, rmin, rmax},

  ScalingFunctions → {If [xscale, "Log", None],
    If[yscale, "Log", None]},
  MeshFunctions → {#3 &}, Mesh → {{0.2, 0.4, 0.6, 0.8}},
  PlotPoints → ControlActive[10, 30],

  RegionFunction →
   Function[{m, r,
    z}, (10. ˄(-0.11408792224566819` + 0.27851883673695` x—
      0.01997874049680844` x^2-0.002490304269884624` x^3 +
      0.00007525048500183394` x^4-0.00007162041164677924` x^5-
0.00003393158521958243` x^6 +
      8.589995554646332`*^−7 x^7 +
      1.132375249329131`*^-6 x^8 +
      2.2299345660512832`*^−8 x^9-1.0475165171649914`*^−8 x^10) /.
x → Log10[m]) <
    r < (10. ˄(0.020013868549526272` +
      0.29811170324848235` x—
    0.02012734730157388` x^2-0.0052918215948260265` x^3-
0.0003311775031243655` x^4 +
    0.00004856681718363753` x^5-0.00001245509278944841` x^6-
1.3074832660503483`*^-6 x^7 +
      8.211419885278952`*^−7 x^8 +
      3.47368749025812`*^−8 x^9-1.1251826465596989`*^−8 x^10) /.
    x → Log10[m])],
  BoundaryStyle → None,
  ColorFunction → (Opacity[1., Blend[{Gray, White}, #]] &),
  ColorFunctionScaling → False,
  Exclusions → None],

(∗Density Plot for Envelope-Silicates Contour Mesh,
  approximated by Power-Series in lg [mass] ∗)

  DensityPlot [(1/(10. ˄(0.020013868549526272` +
      0.29811170324848235` x—
    0.02012734730157388` x^2-0.0052918215948260265` x^3-
0.0003311775031243655` x^4 +
    0.00004856681718363753` x^5-0.00001245509278944841` x^6-
1.3074832660503483`*^-6 x^7 +
      8.211419885278952`*^−7 x^8 +
      3.47368749025812`*^−8 x^9-1.1251826465596989`*^−8 x^10) /.
    x → Log10[m]) -
    1/r)∗m,
  {m, mmin, mmax}, {r, rmin, rmax},

  ScalingFunctions → {If [xscale, "Log", None],
    If[yscale,"Log", None]},
  MeshFunctions → {#3 &},
  Mesh → {{0.1, 0.2, 0.3, 0.4, 0.5, 0.6, 0.7, 0.8, 0.9, 1.,
    1.5, 2., 2.5, 3.}},
```

```
    PlotPoints → ControlActive[10, 50],                                        4.4044933682275176`*^-8 x^10) /.
                                                                            x → Log10[
                                                                              m]), (10.^(0.020013868549526272` +
    RegionFunction →                                                             0.29811170324848235` x—0.02012734730157388` x^2-
     Function[{m, r,                                                          0.0052918215948260265` x^3-0.0003311775031243655` x^4 +
      z}, (10. ^(0.020013868549526272` +                                        0.00004856681718363753` x^5-0.00001245509278944841` x^6-
        0.29811170324848235` x—                                               1.3074832660503483`*^-6 x^7 +
        0.02012734730157388` x^2-0.0052918215948260265` x^3-                    8.211419885278952`*^-7 x^8 +
        0.0003311775031243655` x^4 +                                            3.47368749025812`*^-8 x^9-1.1251826465596989`*^-8 x^10) /.
        0.00004856681718363753` x^5-0.00001245509278944841` x^6-               x → Log10[m]),
        1.3074832660503483`*^-6 x^7 +                                         (10.^(-0.11408792224566819` + 0.27851883673695` x—
        8.211419885278952`*^-7 x^8 +                                            0.01997874049680844` x^2-0.002490304269884624` x^3 +
        3.47368749025812`*^-8 x^9-1.1251826465596989`*^-8 x^10) /.              0.00007525048500183394` x^4-0.00007162041164677924` x^5-
      x → Log10[m]) <                                                           0.00003393158521958243` x^6 +
      r],                                                                       8.589995554646332`*^-7 x^7 +
    BoundaryStyle → None,                                                       1.132375249329131`*^-6 x^8 +
    ColorFunction → (Opacity[1., Blend[{Gray, White}, #]] &),                   2.2299345660512832`*^-8 x^9-1.0475165171649914`*^-8 x^10) /.
    ColorFunctionScaling → False,                                             x → Log10[m])
    Exclusions → None]},                                                    }, {m, mmin, mmax},
                                                                            PlotRange → {rmin, rmax},
env2 == 3,                                                                  ScalingFunctions → {If [xscale, "Log", None],
{(*Density Plot for Envelope-Fe Contour Mesh,                                If[yscale,"Log", None]},
approximated by Power-Series in lg [mass] *)                                PlotStyle → {Directive [Thicknes s[0.007],
                                                                             ColorData [97,"ColorList"] [[1]] ],
    DensityPlot [(1/(10.^(-0.11408792224566819` +                              Directive [Thicknes s[0.007],
        0.27851883673695` x—0.01997874049680844` x^2-                          ColorData [97, "ColorList"] [[2]] ],
        0.002490304269884624` x^3 +                                            Directive [Thicknes s[0.007],
        0.00007525048500183394` x^4-0.00007162041164677924` x^5-                ColorData [97, "ColorList"] [[4]] ] }],
        0.00003393158521958243` x^6 +
        8.589995554646332`*^-7 x^7 +                                        (*Legend for the Temperature Coloring of Planets based on /
        1.132375249329131`*^-6 x^8 +                                        Subscript[T, eq ]with the colorscheme of "Rainbow":*)
        2.2299345660512832`*^-8 x^9-1.0475165171649914`*^-8 x^10) /.
      x → Log10[m]) -                                                           Graphics[
      1/r)*m,                                                                   {
    {m, mmin, mmax}, {r, rmin, rmax},                                          Inset[BarLegend[{"Rainbow", {0, 1}},
                                                                                "Ticks" → {
                                                                                 {Log10[300./300.]/Log10[2000./300.], "300"},
    ScalingFunctions → {If [xscale, "Log", None],                                {Log10[400./300.]/Log10[2000./300.], "400"},
     If[yscale,"Log", None]},                                                    {Log10[500./300.]/Log10[2000./300.], "500"},
    MeshFunctions → {#3 &},                                                      {Log10[700./300.]/Log10[2000./300.], "700"},
    Mesh → {{0.1, 0.2, 0.3, 0.4, 0.5, 0.6, 0.7, 0.8, 0.9, 1.,                     {Log10[1000./300.]/Log10[2000./300.],"1000"},
       1.5, 2., 2.5, 3.}},                                                       {Log10[1400./300.]/Log10[2000./300.], "1400"},
    PlotPoints → ControlActive[10, 50],                                          {Log10[2000./300.]/Log10[2000./300.], "2000"}
                                                                                },
                                                                                "TickSide" → Right,
    RegionFunction →                                                            "TickLengths" → 2,
     Function[{m, r,                                                            "TicksStyle" → Directive [Gray, Thicknes s[0.1]],
      z}, (10. ^(-0.11408792224566819` + 0.27851883673695` x—                   Frame → False,
        0.01997874049680844` x^2-0.002490304269884624` x^3 +                    (*Method /[Rule]{Frame /[Rule]False,
        0.00007525048500183394` x^4-0.00007162041164677924` x^5-                TicksStyle /[Rule]Directive [Black,Thicknes s[0.5]] },*)
        0.00003393158521958243` x^6 +
        8.589995554646332`*^-7 x^7 +                                              LabelStyle →
        1.132375249329131`*^-6 x^8 +                                             Directive [Gray, FontSize → 12, Italic],
        2.2299345660512832`*^-8 x^9-1.0475165171649914`*^-8 x^10) /.              LegendMarkerSize → 90,
      x → Log10[m]) <
      r],                                                                         LegendLabel →
    BoundaryStyle → None,                                                        "!(/*SubscriptBox[/(T/), /(eq/)]/)(K)"],
    ColorFunction → (Opacity[1., Blend[{Gray, White}, #]] &),                   Scaled[{0., 0.47}], {Left, Bottom}],
    ColorFunctionScaling → False,
    Exclusions → None]}, True, {}], {}],                                    (*Legend for Three Pure-Componennt Mass–Radius Curves: Ices,
                                                                            Silicates, and Fe-metals*)
Plot[{(10. ^(0.13666292574887867` + 0.27183702181443314` x—
      0.0071340243326271119` x^2-0.0021407416433092126` x^3-                    Inset[LineLegend[{Directive [Thicknes s[0.007],
      0.0022608931475693915` x^4-0.0002516518649610248` x^5 +                    ColorData [97, "ColorList"] [[1]] ],
      0.00011968169122553435` x^6 +                                              Directive [Thicknes s[0.007],
      0.000011663496987412905` x^7-3.536434693875541`*^-6 x^8-                   ColorData [97, "ColorList"] [[2]] ],
    1.6848230313524644`*^-7 x^9 +                                                Directive [Thicknes s[0.007],
```

```
      ColorData [97, "ColorList"] [[4]] ] }, {"Ices", "Silicates",
    "Fe-metals"},
    LabelStyle → {16, Gray}],
   Scaled[{0., 0.77}], {Left, Bottom}],

(∗Dynamic Annotation within the Figure: ∗)
(∗Mass /
and Radius Errorbars:∗)

  Dynamic[Text[
   Style["/!/(/*SubscriptBox[/(/ [Sigma] M/),/
/(P/)])/)//!/(/*SubscriptBox[/(M/), /(P/)])/)/[LessEqual]" <>
     ToString[NumberForm[/ [Sigma] Mpercent], TraditionalForm] <>
      "
/(P/)])/)//!/(/*SubscriptBox[/(R/), /(P/)])/)/[LessEqual]" <>
     ToString[NumberForm[/ [Sigma] Rpercent], TraditionalForm] <>
      "

(∗Planet Equilibrium Temperature Range:∗)

  Dynamic[Text[
   Style[ToString[NumberForm[First[Teq]], TraditionalForm] <>
     "/[LessEqual]/!/(/*SubscriptBox[/(T/), /(eq/)]/)(K)/
/[LessEqual]" <> ToString[NumberForm[Last[Teq]], TraditionalForm],
    Gray, FontSize → 14], Scaled[{0.99, 0.05}], {1, -1}]],

(∗Host Stellar Radius Range:∗)

  Dynamic[Text[
   Style[ToString[NumberForm[First[rstar]], TraditionalForm] <>
     "/[LessEqual] /!/(/*SubscriptBox[/(R/), /(/[Star/)])/)///
/!/(/*SubscriptBox[/(R/), /(/[Sun/)])/)/[LessEqual]" <>
     ToString[NumberForm[Last[rstar]], TraditionalForm], Gray,
    FontSize → 14], Scaled[{0.98, 0.015}], {1, −1}]],

(∗Solar System Planets Represented as Gray Filled Circles: ∗)

     PointSize[0.033], Gray,
   Point@{If [xscale, Log, Identity]@# [[1]],
    If[yscale, Log, Identity]@# [[2]] } & /@ {{0.1074, 0.53,
     "M"}, {0.815, 0.95, "V"}, {1., 1., "E"}, {14.536, 4.0,
     "U"}, {17.15, 3.883, "N"}, {95.159, 9.14, "S"}, {317.8,
     11.2, "J"}},

(∗Labeling of Solar System Planets with their First Letter: /
∗)

   Text[Style[# [[3]], Bold, Gray,
    FontSize → 12], {If [xscale, Log, Identity]@# [[1]],
    If[yscale, Log, Identity]@# [[
     2]] }, {-1, -2}] &/@ {{0.1074, 0.53, "M"}, {0.815,
     0.95, "V"}, {1., 1., "E"}, {14.536, 4.0, "U"}, {17.15,
     3.883, "N"}, {95.159, 9.14, "S"}, {317.8, 11.2, "J"}},

(∗Interative Rendering of Background Exoplanet Data (from /
TepCat) with Tooltip and Mouseover Functions: ∗)

   PointSize[0.025],
   Dynamic[{ColorData ["Rainbow"][(Log[# [[39]] /300])/
     Log[2000/300]],
    Mouseover[
     Tooltip[
      Point@{If [xscale, Log, Identity]@(# [[27]] ∗317.8),
       If[yscale, Log, Identity]@(# [[30]] ∗11.2)}, # [[
       1]] ], {Line@{{If [xscale, Log,
        Identity]@((# [[27]] - # [[29]] )∗317.8),
       If[yscale, Log, Identity]@(# [[30]] ∗11.2)}, {If[

      xscale, Log,
      Identity]@((# [[27]] + # [[28]] )∗317.8),
      If[yscale, Log, Identity]@(# [[30]] ∗11.2)}},
     Line@{{If [xscale, Log, Identity]@(# [[27]] ∗317.8),
       If[yscale, Log,
        Identity]@((# [[30]] - # [[32]] )∗11.2)}, {If [xscale,
        Log, Identity]@(# [[27]] ∗317.8),
       If[yscale, Log,
        Identity]@((# [[30]] + # [[31]] )∗11.2)}}}]} & /@
   Select [data0 [[2 ;;]],
    mmin ⇐ (# [[27]] ∗317.8) ⇐ mmax && NumberQ [# [[28]] ] &&
     NumberQ [# [[29]] ] && rmin ⇐ (# [[30]] )∗11.2) ⇐ rmax &&
     NumberQ [# [[31]] ] &&
     NumberQ [# [[32]] ] && # [[31]] /# [[30]] ⇐ / [Sigma] Rpercent/
      100. && # [[32]] /# [[30]] ⇐ / [Sigma] Rpercent/
      100. && # [[28]] /# [[27]] ⇐ / [Sigma] Mpercent/
      100. && # [[29]] /# [[27]] ⇐ / [Sigma] Mpercent/
      100. && # [[30]] ˆ4/# [[27]] > 0.01 &&
     Porb [[1]] ⇐ # [[20]] ⇐ Porb [[2]] &&
     Teq [[1]] ⇐ # [[39]] ⇐ Teq [[2]] &&
     Teff [[1]] ⇐ # [[2]] ⇐ Teff [[2]] &&
     eccentricity [[1]] ⇐ # [[21]] ⇐ eccentricity [[2]] &&
     mstar [[1]] ⇐ # [[8]] ⇐ mstar [[2]] &&
     rstar [[1]] ⇐ # [[11]] ⇐ rstar [[2]] &&
     aorb [[1]] ⇐ # [[24]] ⇐ aorb [[2]] &&
     First[FeHdex] ⇐ # [[5]] ⇐ Last[FeHdex] &] ],

(∗Interative Rendering of Input Exoplanet Data from the /
InputField: ∗)
  (∗Draw the new exoplanets: ∗)
   /
(∗Black Rim∗)
   PointSize[0.052],
   Thicknes s[0.015],
   Dynamic@
   If[add1, {Black, {Point@{If [xscale, Log, Identity]@# [[2]],
       If[yscale, Log, Identity]@# [[5]] }, ,

     Line@{{If [xscale, Log, Identity]@(# [[2]] - # [[4]] ),
       If[yscale, Log, Identity]@# [[5]] }, {If [xscale, Log,
        Identity]@(# [[2]] + # [[3]] ),
       If[yscale, Log, Identity]@# [[5]] }},
     Line@{{If [xscale, Log, Identity]@# [[2]],
       If[yscale, Log, Identity]@(# [[5]] - # [[7]] )}, {If[
        xscale, Log, Identity]@# [[2]],
       If[yscale, Log, Identity]@(# [[5]] + # [[6]] )}}}} & /@
    Select [np2,
     If [filter1,
      mmin ⇐ (# [[2]] ) ⇐ mmax && rmin ⇐ (# [[5]] ) ⇐ rmax &&
       Teq [[1]] ⇐ (# [[8]] ) ⇐ Teq [[2]] &
      mmin ⇐ (# [[2]] ) ⇐ mmax &&
       rmin ⇐ (# [[5]] ) ⇐ rmax &]], {}],

(∗Colored Interior∗)
   PointSize[0.045],
   Thicknes s[0.008],
   Dynamic@
   If[add1, {ColorData ["Rainbow"][(Log[# [[8]] /300)/
      Log[2000/300]], {Tooltip[
       Point@{If [xscale, Log, Identity]@# [[2]],
        If[yscale, Log, Identity]@# [[5]] }, # [[1]],
       TooltipStyle →
        Directive [Bold, 20]], (∗try annotation here,
      but unsuccessful!∗)

      Line@{{If [xscale, Log, Identity]@(# [[2]] - # [[4]] ),
        If[yscale, Log, Identity]@# [[5]] }, {If [xscale, Log,
```

```
      Identity]@(# [[2]] + # [[3]] ),
      If[yscale, Log, Identity]@# [[5]] }},
    Line@{{If [xscale, Log, Identity]@# [[2]],
      If[yscale, Log, Identity]@(# [[5]] - # [[7]] )}, {If[
      xscale, Log, Identity]@# [[2]],
      If[yscale, Log, Identity]@(# [[5]] + # [[6]] )}}}}& /@
    Select [np2,
    If [filter1,
      mmin ⇐ (# [[2]] ) ⇐ mmax && rmin ⇐ (# [[5]] ) ⇐ rmax &&
      Teq [[1]]  ⇐ (# [[8]] ) ⇐ Teq [[2]]  &
      mmin ⇐ (# [[2]] ) ⇐ mmax &&
      rmin ⇐ (# [[5]] ) ⇐ rmax &]], {}],

  (∗Labels New Planets' Name:∗)

  Dynamic@If[
    add1 &&
    add2,{Text[
      Framed[# [[1]], Background → White,
      BaseStyle → {Black, Bold, 15}, FrameMargins → 0],
      Offset[{50, 20}∗
      add3, {If [xscale, Log, Identity]@# [[2]],
      If[yscale, Log, Identity]@# [[5]] }]] } & /@
    Select [np2,
    If [filter1,
      mmin ⇐ (# [[2]] ) ⇐ mmax && rmin ⇐ (# [[5]] ) ⇐ rmax &&
      Teq [[1]]  ⇐ (# [[8]] ) ⇐ Teq [[2]]  &
      mmin ⇐ (# [[2]] ) ⇐ mmax &&
      rmin ⇐ (# [[5]] ) ⇐ rmax &]], {}]

  (∗add in some commentary marks:∗)
  (∗Black,Dotted,
  Line[{{7.4,2.36},{10.4,4.14}}],
  Black, Dotted, Line[{{10.5,0.1},{10.5,5.}}],
  HatchFilling[2/[Pi]/4],EdgeForm[None],
  Triangle[{{4.,2.},{10.5,2.4},{8.5,3.}}],

  HatchFilling[],EdgeForm[None],
  Rotate[Disk[{14.,4.},{4,1/3},{0,
  2/[Pi]}],-1Degree],∗)

  }
  ],
  PlotRangePadding → None,
  ImageSize → {450, 450},
  (∗ImagePadding /[Rule]{{left,right},{bottom,top}}∗)

  ImagePadding → {{50, 0}, {50, 0}},
  Frame → True,
  (∗FrameLabel → {{left,right},{bottom,top}}∗)

  FrameLabel → {{"Planet Radius (/!/(/*SubscriptBox[/(R/), /
/(P/)])/)//!/(/*SubscriptBox[/(R/), /(/[Earth]/)]/))",
    ""}, {"Planet Mass (/!/(/*SubscriptBox[/(M/), /(P/)])/)//!/(/
/*SubscriptBox[/(M/), /(/[Earth]/)]/))", ""}},
  (∗FrameTicks /[Rule]{{Automatic,None},{None,Automatic}},∗)

  FrameStyle → Directive [14, Gray, Thicknes s[0.007]],
  AspectRatio → 1
  ]
,
 plt2 = Dynamic@Histogram[{
  (# [[30]] ∗11.2) & /@

  Select [data0 [[2 ;;]],
    mmin ⇐ (# [[27]] ∗317.8) ⇐ mmax && NumberQ [# [[28]] ] &&
    NumberQ [# [[29]] ] && rmin ⇐ (# [[30]] ∗11.2) ⇐ rmax &&
      NumberQ [# [[31]] ] &&
      NumberQ [# [[32]] ] && # [[31]] /# [[30]] ⇐ / [Sigma] Rpercent/
      100. && # [[32]] /# [[30]] ⇐ / [Sigma] Rpercent/
      100. && # [[28]] /# [[27]] ⇐ / [Sigma] Mpercent/
      100. && # [[29]] /# [[27]] ⇐ / [Sigma] Mpercent/
      100. && # [[30]] ^4/# [[27]] > 0.01 &&
      Porb [[1]] ⇐ # [[20]] ⇐ Porb [[2]] &&
      Teq [[1]]  ⇐ # [[39]] ⇐ Teq [[2]]  &&
      Teff [[1]] ⇐ # [[2]] ⇐ Teff [[2]] &&
      eccentricity [[1]] ⇐ # [[21]] ⇐ eccentricity [[2]] &&
      mstar [[1]] ⇐ # [[8]] ⇐ mstar [[2]] &&
      rstar [[1]] ⇐ # [[11]] ⇐ rstar [[2]] &&
      aorb [[1]] ⇐ # [[24]] ⇐ aorb [[2]] &&
      First[FeHdex] ⇐ # [[5]] ⇐ Last[FeHdex] &]
  ,
  If[
    add1,# [[5]] & /@
    Select [np2,
    If [filter1,
      mmin ⇐ (# [[2]] ) ⇐ mmax && rmin ⇐ (# [[5]] ) ⇐ rmax &&
      Teq [[1]] ⇐ (# [[8]] ) ⇐ Teq [[2]] &
      mmin ⇐ (# [[2]] ) ⇐ mmax &&
      rmin ⇐ (# [[5]] ) ⇐ rmax &]], {}]
  },
  (∗{0.1},∗)
  If[yscale,
  {"Log", {10^
    Range[Log[10, rmin],
    Log[10, rmax],(Log[10, rmax] - Log[10, rmin])/
    histradiusbin]}}, {rmin,
    rmax, (rmax—rmin)/histradiusbin }],
  AspectRatio → 5, BarOrigin → Left,
  ChartLayout → "Stacked",
  PlotRangePadding → None,
  ImageSize → {Automatic, 450},
  (∗ImagePadding /[Rule]{{left,right},{bottom,top}}∗)

  ImagePadding → {{0, 50}, {50, 0}},
  Frame → True,
  (∗FrameLabel → {{left,right},{bottom,top}}∗)

  FrameTicks → All,
  FrameLabel → {{"",
    "Histogram of (/!/(/*SubscriptBox[/(R/), /
/(P/)])/)//!/(/*SubscriptBox[/(R/), /(/[Earth]/)]/))"}, {"counts", ""}},
  FrameStyle → Directive [14, Gray, Thicknes s[0.027]] ] }
},
Alignment → {{Right, Left}, {Bottom, Top}},
Spacings → {0, 0}
],

(∗Localize variables and functions ∗)
(∗data0: the un-modified data /
downloaded from TepCat:∗)
{data0, None},
(∗plt: the combined plot: ∗)
{plt, None},

(∗mass–radius plot: ∗)
{plt1, None},
(∗plt2: histogram of (Subscript[R, p]/Subscript[R, /[Earth]] ): ∗)
/
{plt2, None},
(∗plt3: histogram of (Subscript[M, p]/Subscript[M, /[Earth]] ): ∗)
/
{plt3, None},
(∗plt4: histogram of ((Subscript[R, p]/Subscript[R, /
```

```
/[Earth]] )/(Subscript[M, p]/Subscript[M, /[Earth]] )^(1/4)): ∗)
{plt4,
None},

(∗Control Pane Region: Specift Manipulating variables and their /
range∗)
{{Teff, {2500, 10000},
"/!/(/*SubscriptBox[/(T/), /(/(/[Star]/)/(eff/)/)]/)(K)"}, 2500,
10000, ControlType → IntervalSlider, Method → "Push",
Appearance → {"Labeled", "Paired"}},
{{FeHdex, {-0.5, 0.5}, "[Fe/H]"}, -0.5, 0.5,
ControlType → IntervalSlider, Method → "Push",
Appearance → {"Labeled", "Paired"}},
{{mstar, {0., 2.},
"/!/(/*SubscriptBox[/(M/), /
/(/[Star]/)]/)(/!/(/*SubscriptBox[/(M/), /(/[Sun]/)]/))"}, 0., 2.,
ControlType → IntervalSlider, Method → "Push",
Appearance → {"Labeled", "Paired"}},
{{rstar, {0.6, 2.},
"/!/(/*SubscriptBox[/(R/), /
/(/[Star]/)]/)(/!/(/*SubscriptBox[/(R/), /(/[Sun]/)]/))"}, 0., 2.,
ControlType → IntervalSlider, MinIntervalSize → 0.1,
Method → "Push", Appearance → {"Labeled", "Paired"}},
{{Porb, {0.1, 100.}, "/!/(/*SubscriptBox[{P},{orb}]/)(days)"},
0.1,100,ControlType → IntervalSlider,Method → "Push",
Appearance → {"Labeled","Paired"}},
{{aorb,{0.01,0.5},"/!/(/*SubscriptBox[{a},{orb}]/)(AU)"},
0.01,0.5,ControlType → IntervalSlider,Method → "Push",
Appearance → {"Labeled","Paired"}},
{{eccentricity, {0., 1.}, "eccentricity"}, 0., 1.,
ControlType → IntervalSlider, Method → "Push",
Appearance → {"Labeled", "Paired"}},
{{Teq, {900, 1500}, "/!/(/*SubscriptBox[/(T/), /(eq/)/)](K)"}, 200,
3000, ControlType → IntervalSlider, MinIntervalSize → 50,
Method → "Push", Appearance → {"Labeled", "Paired"}},

Delimiter,
{{/[Sigma] Mpercent, 50,
"/!/(/*SubscriptBox[/(/[Sigma] M/), /(P/)]/)/!/(/*SubscriptBox[/(M/
/), /(P/)]/)(
AnimationRate → 1}, {{/[Sigma] Rpercent, 20,
"/!/(/*SubscriptBox[/(/[Sigma] R/), /(P/)]/)/!/(/*SubscriptBox[/(R/
/), /(P/)]/)(
AnimationRate → 1},

Delimiter,

(∗Use FormObject to specify Control and TrackingFunction to enforce /
restrictions on a Control:∗)
(∗Choice of logarithmic-versus-linear /
scale in mass: x-dimension∗)

Grid[{{Control[{{xscale, False,
  "/!/(/*SubscriptBox[{m}, {scale}]/):"}, {False → "linear",
  True → "log"},
  TrackingFunction → (xscale = #;
  If[#, np1 = MapAt[Log, {;;, 1}]@np1,
   np1 = MapAt[Exp, {;;, 1}]@np1];&)}],
(∗min-mass of the plot∗)

Control[{{mmin, 0.1,
  "/!/(/*SubscriptBox[{m}, /
{min}]/)(/!/(/*SubscriptBox[/(M/), /(/[Earth]/)]/))"},
  ControlType → InputField, FieldSize → 5,
  TrackingFunction → (If[NumberQ [#] &&# > 0, mmin = #]&)}],
(∗max-mass of the plot∗)
```

```
Control[{{mmax, 20.,
  "/!/(/*SubscriptBox[{m}, /
{max}]/)(/!/(/*SubscriptBox[/(M/), /(/[Earth]/)]/))"},
  ControlType → InputField, FieldSize → 5,
  TrackingFunction → (If[NumberQ [#] &&# > 0, mmax = #] &)}]},
(∗Choice of logarithmic-versus-linear scale in radius: y-
dimension∗)
{Control[{{yscale, False,
  "/!/(/*SubscriptBox[{r}, {scale}]/):"}, {False → "linear",
  True → "log"},
  TrackingFunction → (yscale = #;
  If[#, np1 = MapAt[Log, {;;, 2}]@np1,
   np1 = MapAt[Exp, {;;, 2}]@np1];&)}],
(∗min-radius of the plot∗)

Control[{{rmin, 0.5,
  "/!/(/*SubscriptBox[{r}, /
{min}]/)(/!/(/*SubscriptBox[/(R/), /(/[Earth]/)]/))"},
  ControlType → InputField, FieldSize → 5,
  TrackingFunction → (If[NumberQ [#] &&# > 0, rmin = #] &)}],
(∗max-radius of the plot∗)
Control[{{rmax, 5.,
  "/!/(/*SubscriptBox[{r}, /
{max}]/)(/!/(/*SubscriptBox[/(R/), /(/[Earth]/)]/))"},
  ControlType → InputField, FieldSize → 5,
  TrackingFunction → (If[NumberQ [#] &&# > 0, rmax = #] &)}]}}],

Delimiter,

(∗manipulate histogram binning:∗)

Row[{"Histogram:",
  Control[{{histmassbin, 11, " mass bin"}, {1, 2, 3, 4, 5, 6, 7, 8,
  9, 10, 11, 12, 13, 14, 15, 16, 17, 18, 19, 20, 21, 22, 23, 24,
  25, 30, 40, 50, 60, 80, 100}, ControlType → PopupMenu}],
  Control[{{histradiusbin, 23, " radius bin"}, {1, 2, 3, 4, 5, 6, 7,
  8, 9, 10, 11, 12, 13, 14, 15, 16, 17, 18, 19, 20, 21, 22, 23,
  24, 25, 30, 40, 50, 60, 80, 100}, ControlType → PopupMenu}],
Control[{{histzetabin, 10, " /[Zeta] bin"}, {1, 2, 3, 4, 5, 6, 7,
  8, 9, 10, 11, 12, 13, 14, 15, 16, 17, 18, 19, 20, 21, 22, 23,
  24, 25, 30, 40, 50, 60, 80, 100}, ControlType → PopupMenu}]
  }],

Delimiter,

(∗Ask: add envelope or not?∗)

Row[{Control[{{env1, True, "add envelope?"}, {False, True}}],
 Control[{{env2, 1,
  " onto"}, {1 → "/!/(/*SubscriptBox[{H}, 2]/)O",
  2 → "Silicates", 3 → "Fe"}}]}],
Item["/!/(/*UnderscriptBox[/(/[Integral]/), /
/(/(/[Element]/)/[env]/)]/)/!/(/*FractionBox[{dP}, /(/[Rho]/)]/)/(/
/!/(/*FractionBox[SubscriptBox[{GM}, /(/[Earth]/)], /
SubscriptBox[/(R/), /
/(/[Earth]/)]] /)) = {0.1,0.2,0.3,0.4,0.5,0.6,0.7,0.8,0.9,1,1.5,2,2.5,3} /
contours"],
Delimiter,
(∗add new planet(s)?∗)
(∗Step1: Ask add planet(s) or not?∗)
/
(∗Step2: Specify the number of planet(s)∗)
Row[{
 Column[
 {
 Control[{{add1, False, "add new planet(s)?"}, {False, True}}],
```

(Continued)

```
 Control[{{filter1, True,
   "filter new planet(s) n by /!/(/*SubscriptBox[/(T/), /
/(eq/)]/)?"}, {False, True}}]
 }
],

Dynamic@
Control[{{add2, True,
  " add new planet(s) n name label(s)?"}, {False, True},
  Enabled → Dynamic[add1]}],
Dynamic@Control[{{add3, {-1, 0.8},
  " label nplacement nslider2D:"}, {-1, -1}, {1, 1},
  ControlType → Slider2D, Enabled → Dynamic[add1 && add2]}]}],
(*Step3: Create InputFields for mass, radius, their errors, and /
temperature for each new planet*)

Item["Input {{planetname, m(/!/(/*SubscriptBox[/(M/), /
/(/[Earth/)]/)),/ [Sigma] m+,/ [Sigma] m-,r(/!/(/*SubscriptBox[/(R/), /(/
/[Earth/)]/)),/ [Sigma] r+,/ [Sigma] r-,/!/(/*SubscriptBox[/(T/), /
/(eq/)]/)(K)},{…}} list:"],
(*np2 stores the data for the new planet(s)*)

Control[{{np2, {{"K2-79b", 11.8, 3.6, 3.6, 4.09, 0.17, 0.12,
  1021}, {"K2-222b", 8.0, 1.8, 1.8, 2.35, 0.08, 0.07, 878}},
  "new nplanets' n{m,r,{…}} list:"}, ControlType → InputField,
  FieldSize → Large, Enabled → Dynamic[add1]}],
Delimiter,
(*Exportplots in various formats (PDF,EPS,JPG) and Planet Data in /
the Selected Range*)
Column[{
 Row[{"Export mass–radius plot:",
  Button["pdf", Export["planetplot.pdf", plt1], Method → "Queued",
   ImageSize → Automatic],
  Button["eps", Export["planetplot.eps", plt1], Method → "Queued",
   ImageSize → Automatic],
  Button["jpg", Export["planetplot.jpg", plt1], Method → "Queued",
   ImageSize → Automatic]
 }],
 Row[{"Export planet table:",
  Button["selected planets",
   Export["planettableselected.txt",
    Prepend[Select [data0 [[2 ;;]],
    mmin ⇐ (# [[27]] *317.8) ⇐ mmax && NumberQ [# [[28]] ] &&
    NumberQ [# [[29]] ] && rmin ⇐ (# [[30]] *11.2) ⇐ rmax &&
    NumberQ [# [[31]] ] &&
    NumberQ [# [[32]] ] && # [[31]]/# [[30]] ⇐ / [Sigma] Rpercent/
    100. && # [[32]]/# [[30]] ⇐ / [Sigma] Rpercent/
    100. && # [[28]]/# [[27]] ⇐ / [Sigma] Mpercent/
    100. && # [[29]]/# [[27]] ⇐ / [Sigma] Mpercent/
    100. && # [[30]]  4/# [[27]] > 0.01 &&
    Porb [[1]] ⇐ # [[20]] ⇐ Porb [[2]] &&
    Teq [[1]] ⇐ # [[39]] ⇐ Teq [[2]] &&
    Teff [[1]] ⇐ # [[2]] ⇐ Teff [[2]] &&
    eccentricity [[1]] ⇐ # [[21]] ⇐ eccentricity [[2]] &&
    mstar [[1]] ⇐ # [[8]] ⇐ mstar [[2]] &&
    rstar [[1]] ⇐ # [[11]] ⇐ rstar [[2]] &&
    aorb [[1]] ⇐ # [[24]] ⇐ aorb [[2]] &&    First
    [FeHdex] ⇐ # [[5]] ⇐ Last[FeHdex] &], data0 [[1]] ],
    "Table"], Method → "Queued", ImageSize → Automatic],
  Button["all planets",
   Export["planettableall.txt", data0, "Table"],
   Method → "Queued", ImageSize → Automatic]}],
 Row[{"Export histogram of planet radius:",
  Button["pdf", Export["histogram.pdf", plt2], Method → "Queued",
   ImageSize → Automatic],
  Button["eps", Export["histogram.eps", plt2], Method → "Queued",
   ImageSize → Automatic],
```

(Continued)

```
  Button["jpg", Export["histogram.jpg", plt2], Method → "Queued",
   ImageSize → Automatic]}],
 Row[{"Export histogram of planet mass:",
  Button["pdf", Export["ternaryplot.pdf", plt3],
   Method → "Queued", ImageSize → Automatic],
  Button["eps", Export["ternaryplot.eps", plt3],
   Method → "Queued", ImageSize → Automatic],
  Button["jpg", Export["ternaryplot.jpg", plt3],
   Method → "Queued", ImageSize → Automatic]}],
 Row[{"Export histogram of /[Zeta]:",
  Button["pdf", Export["histogram.pdf", plt4], Method → "Queued",
   ImageSize → Automatic],
  Button["eps", Export["histogram.eps", plt4], Method → "Queued",
   ImageSize → Automatic],
  Button["jpg", Export["histogram.jpg", plt4], Method → "Queued",
   ImageSize → Automatic]}],
 Row[{"Export combined plot:",
  Button["pdf", Export["combinedplot.pdf", plt],
   Method → "Queued", ImageSize → Automatic],
  Button["eps", Export["combinedplot.eps", plt],
   Method → "Queued", ImageSize → Automatic],
  Button["jpg", Export["combinedplot.jpg", plt],
   Method → "Queued", ImageSize → Automatic]}]
}]
,

ControlPlacement → Left,
SynchronousUpdating → False,
ContinuousAction → False,
SaveDefinitions → True,
Deployed → True,
(*Load external exoplanet data from TepCat compiled by John /
Southworth from UK: *)

Initialization :> ({data0 = Import["https://www.astro.keele.ac.uk/jkt/tepcat/
  allplanets-/
  ascii.txt","Table"]})]
```

## ORCID iDs

Li Zeng  https://orcid.org/0000-0003-1957-6635
Stein B. Jacobsen  https://orcid.org/0000-0003-2164-0720
Eugenia Hyung  https://orcid.org/0000-0001-6635-4918
Amit Levi  https://orcid.org/0000-0003-1080-1007
Chantanelle Nava  https://orcid.org/0000-0001-8838-3883
James Kirk  https://orcid.org/0000-0002-4207-6615
Caroline Piaulet  https://orcid.org/0000-0002-2875-917X
Gaia Lacedelli  https://orcid.org/0000-0002-4197-7374
Dimitar D. Sasselov  https://orcid.org/0000-0001-7014-1771
Sarah T. Stewart  https://orcid.org/0000-0001-9606-1593
Munazza K. Alam  https://orcid.org/0000-0003-4157-832X
Mercedes López-Morales  https://orcid.org/0000-0003-3204-8183
Mario Damasso  https://orcid.org/0000-0001-9984-4278
David W. Latham  https://orcid.org/0000-0001-9911-7388